\documentclass[preprint]{aastex}

\usepackage{emulateapj5}

\def\gsim{\;\lower.6ex\hbox{$\sim$}\kern-7.75pt\raise.65ex\hbox{$>$}\;}
\def\lsim{\;\lower.6ex\hbox{$\sim$}\kern-7.75pt\raise.65ex\hbox{$<$}\;}
\def\aa{A\&A }
\def\mv{$m_{F555W}$}
\def\mi{$m_{F814W}$}
\def\mj{$m_{F110W}$}
\def\mh{$m_{F160W}$}
\def\mvi{$m_{F555W}-m_{F814W}$}
\def\mjh{$m_{F110W}-m_{F160W}$}

\newcommand{\MSUN}{$M_{\odot}$}
\newcommand{\Myr}{M$_{\odot}$~yr$^{-1}$}

\begin{document}
\title{The Star Formation History of NGC~1705: a Post--Starburst Galaxy on the Verge of Activity\footnotemark[6]}
%\footnote{Based on observations with the NASA/ESA Hubble
%Space Telescope, obtained at the Space Telescope Science Institute,
%which is operated by AURA for NASA under contract NAS5-26555}} 

\author{F. Annibali$^{1,2}$, L. Greggio$^{1,3}$, M. Tosi$^1$, A. Aloisi$^4$, 
Claus Leitherer$^5$}

\affil{$^1$ INAF-Osservatorio Astronomico di Bologna, Via Ranzani 1,
       I-40127 Bologna, Italy\\
       e-mail: greggio@bo.astro.it, tosi@bo.astro.it}

\affil{$^2$ SISSA, via Beirut 4, 34014 Trieste, Italy\\
       e-mail: annibali@sissa.it}

\affil{$^3$ INAF-Osservatorio di Padova, Vicolo dell'Osservatorio 5,
       I-35122 Padova, Italy \\
       e-mail: greggio@pd.astro.it}

\affil{$^4$ Johns Hopkins University
       3400 North Charles St., Baltimore, MD 21218\\
       e-mail: aloisi@pha.jhu.edu}

\affil{$^5$ Space Telescope Science Institute, 
       3700 San Martin Drive, Baltimore, MD 21218\\
       e-mail: leitherer@stsci.edu}

\affil{$^6$Based on observations with the NASA/ESA Hubble
Space Telescope, obtained at the Space Telescope Science Institute,
which is operated by AURA for NASA under contract NAS5-26555}

\begin{abstract}

We infer the star formation history in different regions of the blue compact dwarf NGC~1705 by comparing synthetic color-magnitude diagrams with HST optical and near-infrared photometry. We find that NGC~1705 is not a young galaxy because its star formation commenced at least 5 Gyr ago. On the other hand, we confirm the existence of a recent burst of star formation between 15 and 10 Myr ago. We also find evidence for new strong activity, which started 3 Myr ago and is still continuing. The old population is spread across the entire galaxy, while the young and intermediate stars are more concentrated in the central regions. We derive an almost continuous star formation with variable rate, and exclude the presence of long quiescent phases between the episodes during the last $\approx$ 1 Gyr. The central regions experienced an episode of star formation of $\sim 0.07$ \MSUN $yr^{-1}$ (for a Salpeter initial mass function [IMF]) 15 to 10 Myr ago. This coincides with the strong activity in the central super star cluster. We find a rate of $\sim 0.3$ \MSUN $yr^{-1}$ for the youngest ongoing burst which started $\sim$ 3 Myr ago. This is higher than in other dwarfs and comparable to the rate of NGC~1569. The star formation rate of earlier episodes is not especially high and falls in the range $10^{-3}-10^{-1}$ \MSUN $yr^{-1}$. The IMF is close to the Salpeter value or slightly steeper. 

\end{abstract}


%\clearpage 

\section{Introduction}

Late type dwarf galaxies are important for cosmological issues. With their high gas content and low metallicity, they can be regarded as analogues to primeval galaxies (e.g. Izotov \& Thuan 1999). They are the preferred sites to determine the primordial $^4$He abundance (e.g. Izotov, Thuan \& Lipovetsky 1997; Olive, Skillman \& Steigman 1997; Peimbert, Peimbert \& Ruiz 2000). In hierarchical models for structure formation, dwarfs are the basic building blocks for the assembly of other galaxy types. Besides, dwarf irregulars and blue compact dwarfs (DIrrs and BCDs) could be the local counterparts of the blue population at intermediate redshift. This population is overabundant in deep galaxy counts compared with predictions by models without evolution (e.g., Lilly et al. 1995; Babul \& Ferguson 1996, hereafter BF).

To address the last two issues, the quantitative determination of the star formation history (SFH) in these galaxies is of fundamental importance. Several studies have attempted to decode the information in observed color-magnitude diagrams (CMDs) in terms of stellar ages and star-formation intensities with the synthetic CMD method (Tosi et al. 1991, hereafter T91; Greggio et al. 1998). This method has allowed numerous groups to infer the SF histories of these galaxies with unprecedented accuracy (e.g., T91; Aparicio et al. 1996; Tolstoy \& Saha 1996; Grebel 1998). 

The main results of these studies are:

\noindent$\bullet$ In the Local Group, DIrrs have a {\it gasping} rather than {\it bursting}, SFH, i.e., long periods of moderate activity, separated by short quiescent phases (T91; Greggio et al. 1993; Grebel 1998).

\noindent$\bullet$ Similar to DIrrs, nearby BCDs show no evidence of extended gaps in their SFH (Schulte-Ladbeck et al. 2001).

\noindent$\bullet$ So far, the only case of a (relatively) high SF rate (SFR $\simeq$ 1 \Myr) has been documented for the central region of NGC~1569 (Greggio et al. 1998). This SFR is required to make such galaxies contribute to the faint galaxy counts at the appropriate redshift (BF).

\noindent$\bullet$ Dwarfs in their very first SF episode have remained elusive. The best candidate is IZw18, which hosts stars as old as $\simeq$ 0.3 Gyr or more (Aloisi, Tosi, \& Greggio 1999; \"Ostlin 2000; Hunt, Thuan, \& Izotov 2003).

However, the number of galaxies thoroughly investigated is still small, and more objects must be studied to understand the general picture of the SFH in gas rich dwarfs. In this paper we present the results of the analysis of the CMDs obtained from HST images of the nearby, post-starburst BCD NGC~1705, whose characteristics are similar to that of the exceptionally active dwarf NGC~1569. 

NGC~1705 (M$_{B}$ = -16, $(B-V)$=0.38) shows the best observational evidence of gas outflows (galactic winds) triggered by SN explosions (Meurer et al. 1992, hereafter MFDC; Heckman et al. 2001). Its nuclear region hosts an extremely luminous super star cluster (SSC), which is responsible for $\sim$ 50 $\%$ of the total UV light from the galaxy (MFDC; Heckman \& Leitherer 1997). From the analysis of an HST UV spectrum, Heckman \& Leitherer (1997) inferred an SSC age of at least $\sim$ 10 Myr from the lack of strong O star wind profiles. The estimated mass of the SSC is $\sim 10^5$ \MSUN (Melnick, Moles \& Terlevich 1985, O'Connell, Gallagher \& Hunter 1994; Ho \& Filippenko 1996). Sternberg (1998) found the SSC in NGC~1705 to be overluminous for its mass (and age) and ascribed this to a top-heavy IMF. 

 NGC~1705 is considered a {\it post-starburst} galaxy. The main reason for this classification is the lack of O and W-R star features in the spectra of the SSC. Some scattered, continuing SF does occur in the galaxy: MFDC identified five candidate HII regions and one extended ionizing association ({\it g+h} in their nomenclature). This activity is localized in the high surface brightness (HSB) portion, where the IUE spectrum reveals W-R emission-lines. Since W-R stars have ages $\lesssim 5$ Myr, MFDC suggested ongoing SF during the past 5 Myr in the HSB region, as opposed to the absence of SF in the SSC. The gas metallicity in NGC~1705 has been derived from UV, optical and NIR spectra of HII regions (Storchi-Bergmann, Calzetti \& Kinney 1994): the derived oxygen abundance is 12+log(O/H) = 8.36 and corresponds to Z$\simeq$0.004, i.e., close to the metallicity of the SMC.

\begin{figure}
\figurenum{1}
\epsscale{1.12}
\plotone{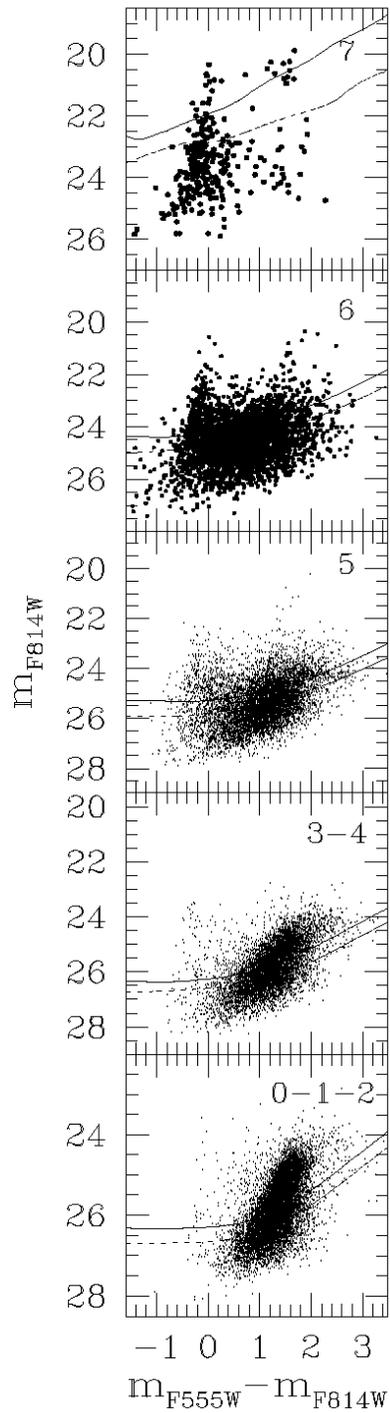}
\label{5obs}
\caption{Optical observed CMDs without the 
%The optical observed CMDs plotted without the
 {\it pearl necklace} objects and with the same color-magnitudes limits
 of the synthetic CMDs in Figs.~\ref{col_sim7},~\ref{col_sim6},
 ~\ref{col_sim5},~\ref{col_sim34} and ~\ref{col_sim012}. 
 %Regions 0, 1 and 2 are plotted together
 %since we have treated them as a whole in the simulations.
 We overplot on the data the completeness levels at 75 
 \%  (solid line) and 50 \% (dotted line). }
\end{figure}

Tosi et al. (2001, hereafter T01) obtained deep HST-WFPC2 photometry in the 
F555W and F814W bands of almost the entire galaxy and divided it into 8 
separate zones (Regions 7 through 0) based on the photometric and density 
characteristics. The optical CMDs presented in T01 are reproduced in 
Fig.~\ref{5obs}, where we have removed the bright objects defined by T01 as
{\it pearl necklace} around the SSC, because they might turn out to be small,
unresolved clusters or associations. 
We have overplotted on the data the 
completeness levels at 75 \% and 50 \%.
The five panels in this figure correspond to the 
different concentric regions of the galaxy, from the center (Region 7) to the 
outermost part (Region 0). 

 From their analysis T01 concluded that:
\begin{itemize}
\item The most massive stars, younger than 10-20 Myr, are mostly concentrated
within 100 pc from the galactic center and the SSC.

\item Intermediate-mass stars with ages up to 1 Gyr are visible only in 
Regions 6 and 5, i.e. within $\sim$ 500 pc from the center. Intermediate-age 
stars are probably present in Region 7 as well, but the extreme crowding 
and high background of this zone makes their detection highly improbable.

\item Beyond $\sim$ 500 pc from the center, the galaxy is mostly populated
by low mass stars on the RGB, with ages from a few Gyr to a Hubble time.

\end{itemize}

T01 data nicely confirm the conclusions of MFDC, which were based on the analysis of the integrated light from NGC~1705.

%with the young, blue stars concentrated towards the center, while the 
%outermost zones are populated almost exclusively by red giant branch (RGB) 
%and asymptotic giant branch (AGB) stars. 
The inner regions were also covered 
with NICMOS (NIC2) and the F110W and F160W filters. The detection of the RGB 
tip in the optical CMD allowed  T01 to give a robust estimate of the 
distance to NGC~1705: $(m-M)_0$ = 28.54 $\pm$ 0.26, tightening the 
previous value by O'Connell et al. (1994).

T01 drew some suggestions on the evolutionary status of NGC~1705 by
simply overimposing theoretical stellar evolution tracks on the empirical CMDs.
In this paper, we apply the synthetic CMD method to the diagrams resulting 
from T01's photometry with the aim of deriving the quantitative SFH in NGC~1705.
To do that, we split the galaxy in 5 concentric regions, each characterized
by similar crowding conditions. This is necessary, since the results of
the simulations are very sensitive to the description of the photometric 
errors. We thus derive accurate results on the SFH occurred in different 
locations within the galaxy.
In Section~2 we present the synthetic CMD method and illustrate in 
some detail the simulations computed for the different regions, so as
to show the basic arguments that support our conclusions.
Readers interested only in the final results may skip this section, and
go directly to Section~3, where we present a summary of our derived
SFH as a function of location in the galaxy. 
In Section 4 we discuss our results and compare them with the SFHs 
inferred in other dwarfs.

\section{The synthetic CMD method}

The most direct way of estimating epochs and intensities of the SF activity in a galaxy is a CMD analysis. 

Synthetic CMDs are constructed via Monte Carlo extractions of (mass, age) pairs for an assumed IMF, SF law, and initial and final epochs of the SF activity. We place each synthetic star in the theoretical $log(L/L_{\odot}), log(T_{eff})$ plane via interpolation on evolutionary tracks. Luminosity and effective temperature are transformed into the desired photometric system by interpolating in bolometric correction tables. The resulting absolute magnitude is converted to a {\it provisional} apparent magnitude by applying the (either known or arbitrary) reddening and distance modulus. Then, an incompleteness test is performed, based on  artificial star tests on the observed frames. This test provides the fraction of recovered stars as a function of input magnitude in each photometric band. When constructing the optical (IR) CMDs, the retained star has to pass both \mv \ and \mi \ (\mj, \ \mh) tests. We assign photometric errors as derived from the distribution of the (output$-$input) mags of the artificial stars (with input mag equal to the {\it provisional} apparent magnitude). These errors take into account both the various instrumental and observational effects, but also systematic uncertainties due to crowding (i.e., the blend of fainter objects into an apparently brighter one). When the number of objects populating the whole synthetic CMD (or a specific portion) equals that of the observed one, the procedure is stopped. We can then derive the SF rate that is consistent with the observations for the prescribed parameters: IMF slope, shape and epochs of the SF, metallicity, distance, and reddening.

The solution provided by the method is not unique, given the photometric and systematic uncertainties, as well as intrinsic degeneracies of stellar evolution (which bring stars of different mass, age, and composition to the same location on the CMD). However, we can significantly reduce the range of possible solutions for the evolutionary status of the examined region.

\subsection{Application to NGC~1705}

\begin{figure}
\figurenum{2}
\epsscale{}
\plotone{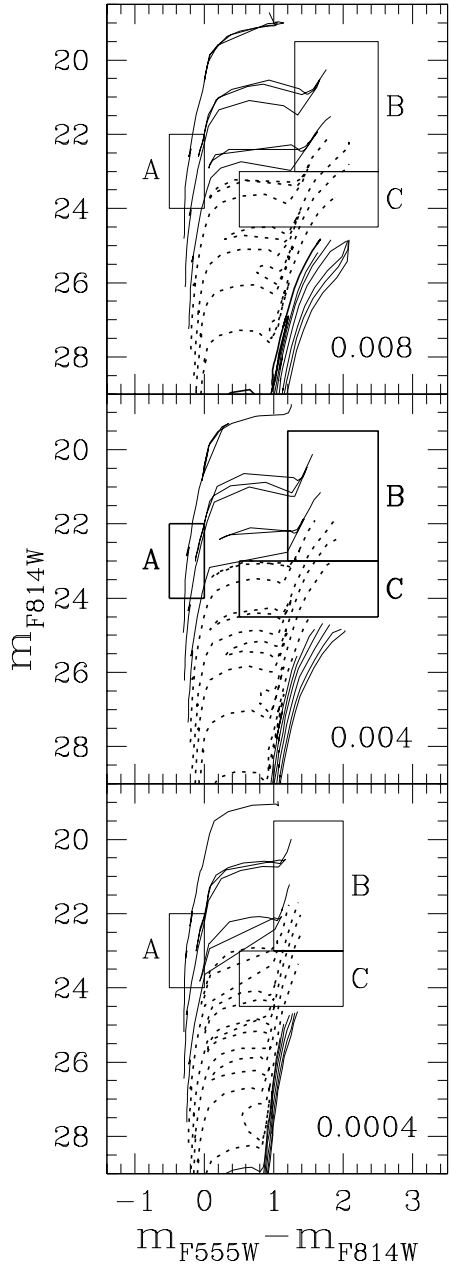}
\label{tracks}
\caption{The Padova stellar evolution tracks (Fagotto et al. 1994a, 1994 b).   
The tracks in the top panel have the metallicity $Z=0.008$, those in the       
middle one have $Z=0.004$, and those in the bottom panel have $Z=0.0004$.      
We plot only the following stellar masses (left to right): 30, 15, 9,    
7, 5, 4, 3, 2, 1.8, 1.6, 1.4, 1.2, 1.0 and 0.9 \MSUN. The corresponding        
lifetimes (which differ slightly from one metallicity to the other) are 7,     
15, 35, 56, 110, 180, 370, 1120, 1300, 2520, 4260, 8410 and 12500 Myr,  
respectively. The boxes plotted on the tracks map different
stellar ages: last 10 Myr in box A, last 60-10 Myr in box B,
ages up to 1 Gyr ago in box C.
At magnitudes fainter than \mi $\lsim$ 24.5 the CMD is populated with 
core helium burning objects older than 100 Myr, plus RGB stars, in principle 
as old as the Hubble time.}
\end{figure}

We adopted the evolutionary tracks by the Padova group (Fagotto et al. 1994a; 1994b) and the conversion tables into the HST-Vega-magnitude system by Origlia \& Leitherer (2000). We have considered sets with metallicity Z=0.0004, 0.004 and 0.008. This metallicity range brackets the appropriate value of NGC~1705. Following the results in T01, we adopt a distance modulus of $(m-M)_{\circ}=28.54$ and a reddening of $E(B-V)=0.045$. The theoretical tracks after transformation into the observational \mi, \mvi \ plane are plotted in Fig.~\ref{tracks}.

 In our models, we start with episodes at constant SF in order to account for the total number of detected objects. This explorative case allows us to study the effects of the photometric errors and incompleteness on the stellar distribution, and to find which regions of the CMD (and thus which ages) are misrepresented in the constant SFR solution. The effects of different IMFs \footnote{Throughout this paper we parameterize the IMF as $\phi (m) \propto m^{-\alpha}$ within the mass range 120-0.1 \MSUN. In this notation, a Salpeter slope has $\alpha=2.35$.} are also tested by treating the slope as a free parameter.  When a constant SF rate is not adequate to describe the data, we model separately the episodes of SF, which populate specific regions of the CMD (hereafter indicated as the age-box technique). In Fig.~\ref{tracks} we schematically show the regions 
selected to constrain the different SF episodes. The location of the age-boxes
depends on the metallicity of the tracks. 
For $Z=0.004$ (corresponding to the HII region metallicity):

\begin{itemize}

\item The blue box at $22<$\mi$<24$, $-0.5<$\mvi$<0$ intercepts very massive stars: it maps the SF over the last $\simeq$ 10 Myr (box A).
\item The (\mi $\lsim$ 23, \ \mvi $\gsim$ 1.2) region collects helium-burning, red objects with masses between 7 and 20 \MSUN (more massive tracks do not extend to the red): this region maps the SFH over the last 10--60 Myr (box B).
\item We find red He burners with masses between 5 and 7 \MSUN (ages between 60 and 100 Myr) plus AGB stars with ages of up to 1 Gyr in the box 23 $\lsim$ \mi $\lsim$ 24.5, \ \mvi $\gsim 0.5$ (box C).

\end{itemize}

These locations are mostly populated with stars in the specific age ranges. 
The star counts in each box are directly linked to the mass that went into 
stars in the SF during the corresponding episodes (see Greggio 2002), and thus 
to the average SFR over the specific time intervals. 
We then construct partial models, which are required to match 
the counts in each box (within the statistical uncertainty) 
 with a SF episode at the appropriate ages.  
Each episode populates the age-boxes in other regions of 
the CMD, as well. 
We account for this when we model those regions of the CMD where there is 
 a large age overlap.
For example, at \mi $\gsim$ 24.5 and \mvi $\gsim$ 0.5, the CMD is populated  
with intermediate age, core helium burning and AGB stars plus RGB stars, 
with ages from $\approx$ 1 Gyr up to the Hubble time. 
The contribution of the intermediate age population to the counts is 
fixed from the partial models constrained by box C;
the remaining stars are produced with SF episodes at older epochs. 
In each partial model we also test the age boundaries, and the effect of 
the IMF, when the data sample a sufficiently large mass interval.  

The box limits described above do not take into account the effect of the photometric errors that shift the stellar magnitudes and colors with respect to the theoretical values.  When constructing the simulations, we first run some preliminary cases in order to accurately define the constraining boxes. 

The evolutionary tracks of intermediate mass in our database end at the first thermal pulse, i.e., they do not contain the thermally pulsing AGB (TP-AGB) phase. We exclude the TP-AGB stars on the synthetic CMD for two reasons: 1) their masses and lifetimes are very uncertain, due to the difficulty of modeling the TP-AGB evolutionary phase; 2) their color transformations to the observational plane are also very uncertain, due to the low surface temperatures. This also affects our ability to describe the photometric errors and incompleteness factors, which could be large, given the very red colors. The TP-AGB population appears more numerous in the IR than in the optical CMD, similar to the case of VII Zw 403 (Schulte-Ladbeck et al. 1999b). When forced to reproduce the total counts 
 in box C, the simulation will overestimate the SFR in the intermediate age range because of our exclusion of the TP-AGB phase. Although short, this phase could well be relevant when the SFR at intermediate ages is high. Therefore we also explore cases constrained by the counts in a bluer part of box C (e. g. 0.5 $\lsim$ \mvi $\lsim$ 1.8) to bracket the SFR at intermediate ages.

Given the much higher spatial coverage, photometric depth, and resolution of the optical over the NIR data, we first performed the simulations in the \mi, \mvi \ plane. The derived SFH was then used as an input to construct the synthetic NIR CMD for the inner regions of the galaxy, to check for consistency. Annibali et al. (2001) presented preliminary results on the SF history in the NIC2 field. 

As stated, the {\it pearl necklace} objects of T01 are excluded. The total number of objects considered in the optical CMD is 282, 3010, 8429, 8513, 11111 from the central Region 7 to the outermost Regions 0$-$1$-$2, respectively. The corresponding CMDs are shown in Fig.~\ref{5obs}. They are plotted on the same scale as the simulations to facilitate the comparison. Notice the different magnitude range in the panels. There are 65, 1079, and 1137 stars in Region 7, 6, and 5, respectively, of the NIR CMD. In the following we describe the results for each region. We will use the following color code: blue dots for ages $<$ 10 Myr, cyan for 10 Myr$<$age$<$60 Myr, green for 60 Myr$<$age$<$1 Gyr; for ages $>$ 1 Gyr, we will use red and black dots for the two metallicities  $Z=0.004$ and $Z=0.0004$, respectively.

\subsubsection{Region 7}

The innermost region has a well-defined blue plume (at \mvi $\simeq -0.1$), as well as two groups of red objects (\mvi $\simeq 1.7$), one bright (at \mi $\sim$ 19.8) and one at fainter magnitudes, down to the limit of our photometry. The blue plume extends up to \mi $=20.3$ and includes stars in the main-sequence (MS) phase and at the hot edge of the core helium burning phase. Comparison with the theoretical tracks of Fig.~\ref{tracks} identifies the brightest of the red ``clumps'' as red supergiants (RSGs) with a mass of 15$-$20 \MSUN (i.e. ages of about 10$-$15 Myr). The fainter red objects are intermediate-mass, core helium-burning and AGB stars. The clear separation of the two groups of red stars indicates a relatively low SFR at epochs between the ages of the fainter and the brighter RSGs.

Because of the severe crowding affecting this central region, we cannot reliably detect truely single objects for magnitudes fainter than about 25.5 in both \mi \ and \mv, implying a look-back time not longer than few hundreds of Myr.   This is a conservative estimate: it refers to the age of the faintest blue loop stars sampled. The red part of the CMD is populated by AGB stars whose age could be as high as 600~Myr. 

We have performed several simulations with different SF scenarios and IMF slopes $\alpha$ (values for $\alpha$ in the range between 1.5 and 2.6 have been tested) to derive the SFH in this region. Since only the young population is detected in Region 7, we performed our simulations with the two most metal rich tracks; the $Z=0.004$ set (corresponding to the value derived in the HII regions) turned out to yield stellar colors in better agreement with the data than the $Z=0.008$ set.

\begin{figure}
\figurenum{3}
\epsscale{}
\plotone{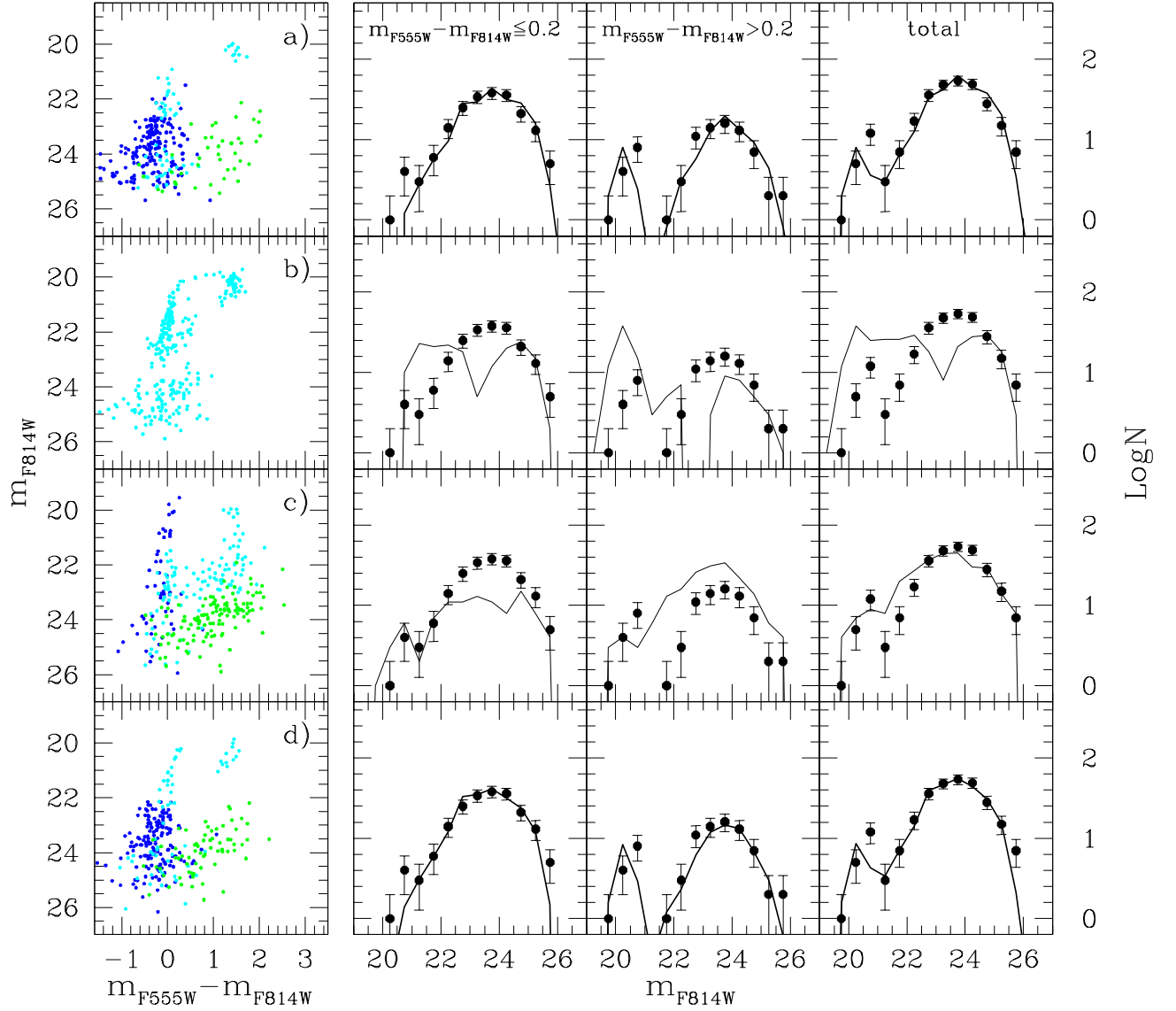}
\label{col_sim7}
\caption{Region 7: synthetic CMDs and LFs. 
 In the synthetic CMDs, different colors correspond to different stellar
 ages: blue for t $< 10$ Myr, cyan for 10 Myr $\leq$ t $<$ 60 Myr,
 green for 60 Myr $\leq$ t $<$ 1 Gyr.	
 Panel a) represents our best case,
 obtained assuming a three-episode scenario: 
 a young burst starting 2 Myr
 ago and still ongoing, a burst occurred from 15 to 10 Myr ago, and
 an older episode starting  0.6 Gyr ago and terminating 60 Myr ago. 
In panel b) all the 282 objects of the observed CMD are produced in a single
 episode (15-10) Myr ago.
 Panel c) shows the case of a continuous SF over the last 0.6 Gyr.
 Case a), b) and c) have been produced with a Salpeter's IMF ($\alpha=2.35$) .
 In panel d) the SFH is the same as in panel a), but $\alpha=2$.
 For all the panels, the adopted metallicity is $Z=0.004$.
 Next to the synthetic CMDs we display the LFs. Dots are for the data, 
 continuous line for the simulations. From left to right, the LFs 
 refer respectively to \mvi $\leq 0.2$,  \mvi$>0.2$ and to the entire 
 color range.}

\end{figure}

Some illustrative examples of the results obtained with $Z=0.004$ are shown in Fig.~\ref{col_sim7}. Plotted are CMDs and LFs for the blue (\mvi $\leq 0.2$) and the red (\mvi $>0.2$) component, plus the total LF. Because of the few objects resolved in this region and the poor statistics, we performed many runs with the same parameters but different random numbers. The derived SFRs are average values. Furthermore, the LFs have been obtained through an average of all the runs for each astrophysical case. The color codes refer to the age range spanned by the synthetic stars.

It is instructive to consider how the CMD is populated by one single burst, which occurred between 10 and 15 Myrs ago (burst 1, hereafter B1). Panel b) in Fig.~\ref{col_sim7} shows this case. The model was obtained with $\alpha$=2.35 and required that all the 282 stars in the observed CMD belong to this episode. This burst, while providing the bright red and blue SGs at the appropriate magnitudes and colors, clearly lacks the faint red giants. In addition, it has a gap in the blue at $23 <$ \mi $< 24$ with no observational counterpart and generates too many bright evolved stars with respect to the MS. This is noticeable both in the CMD and in the LFs. The gap in the blue (commonly called the Hertzsprung gap) results from a truncated SF activity 10 Myr ago, as suggested by the turn-off of the 20 \MSUN track at \mi $\simeq$ 24. The blue plume objects with \mi $<$ 23 are blue loop stars. The discrepancies with respect to the data indicate the need to increase the age range with active SF, both for young and old ages.

At the opposite extreme, the simulation in panel c) of Fig.~\ref{col_sim7} shows the result for a constant SFR over the last 0.6 Gyr (again for a Salpeter IMF). In this case, no gap is apparent in the blue; besides, the blue plume is underpopulated at \mi $>22$ and, except for the brightest bins, there is an excess of red stars over the whole magnitude range. Clearly, we need to introduce a discontinuity in the SF history in order to produce the gap in the distribution of the RGs.  

In panel a) of Fig.~\ref{col_sim7} we present a case in good agreement with the observations. The adopted IMF slope is again Salpeter.  The cyan population is well constrained by its RSG component (11 stars), and results from an episode that occurred between 10 and 15 Myr ago  (B1). We probed different age boundaries and found that a termination more recent than 9 Myr ago would produce too bright SGs; an onset older than 16 Myr ago would populate the gap present in the red. This SF episode accounts for a total of 50 stars (on average) in the observed CMD.   

The fainter group of RGs constrains the SF over older ages. The green population in panel a) corresponds to a SF episode starting 0.6 Gyr ago and terminating 60 Myr ago. If the burst stops less than 30~Myr ago, the red clump of this episode reaches too bright magnitudes and starts populating the gap. The onset of this episode is not well constrained and could be as young as 0.1 Gyr because this portion of the CMD samples stars of quite different mass. Ages older than 0.6 Gyr cannot be probed in Region 7.

The sum of these two episodes accounts for a total of $\sim$ 100 objects in the CMD, falling short of $\sim$ 180 blue plume stars with respect to the data. This third component has been modeled with a very recent burst, (burst 2, hereafter B2) starting 2 Myr ago and still continuing. We find that the onset of this episode needs to be less than 3 Myr ago, otherwise the blue plume reaches too bright magnitudes (see e.g. the blue population in panel c) of Fig.~\ref{col_sim7}). Burst B1 already accounts for the number of stars counted in the brightest bins of the blue plume. 

The SFRs derived for the three episodes depend on the age boundaries: when decreasing the burst duration, the SFR increases. Tab.~\ref{tab_1} shows the values of the SFR in the three episodes obtained by varying the age boundaries within the range allowed by the data. We notice that the current SFR is quite high: to make it lower, we should let the episode last longer. This however would lead to an overproduction of bright blue stars. It seems that the SFR has been increasing from the past towards the most recent epochs.

The SFR values in Tab.~\ref{tab_1} have been obtained by extrapolating the IMF down to 0.1 \MSUN, with a constant $\alpha$. These extrapolated values are very sensitive to the IMF slope. In our CMD, only the two younger episodes probe a mass range large enough to be sensitive to the effects of a different IMF. We find that an IMF as flat as $\alpha=2$ still produces CMDs in agreement with the observations. The CMD in panel d) results from the same three-episode scenario of panel a), but $\alpha=2$. An even flatter IMF produces a too flat LF for the blue plume. On the other hand, IMFs steeper than Salpeter yield too steep LFs for the youngest burst. The SFR values for the $\alpha=2$ models are quoted in Tab.~\ref{tab_1}. Compared to the Salpeter case, the temporal behavior of the SFR does not change, but the levels are systematically lower.

\subsubsection{Region 6}

In Fig.~\ref{5obs} we show the observed \mi, \mvi \ diagram for Region 6. As in Region 7, one can recognize well-defined blue and red plumes. The dramatic drop of crowding away from the central region containing the SSC allows the detection of much fainter objects. We notice the concentration of objects with $1\leq$ \mvi $\leq 1.8$ and $24.2 \leq$ \mi $\leq 26$, (better delineated from Region 5 outwards), which likely   correspond to low-mass, old stars in the RGB phase.

\begin{figure}
\figurenum{4}
\epsscale{}
\plotone{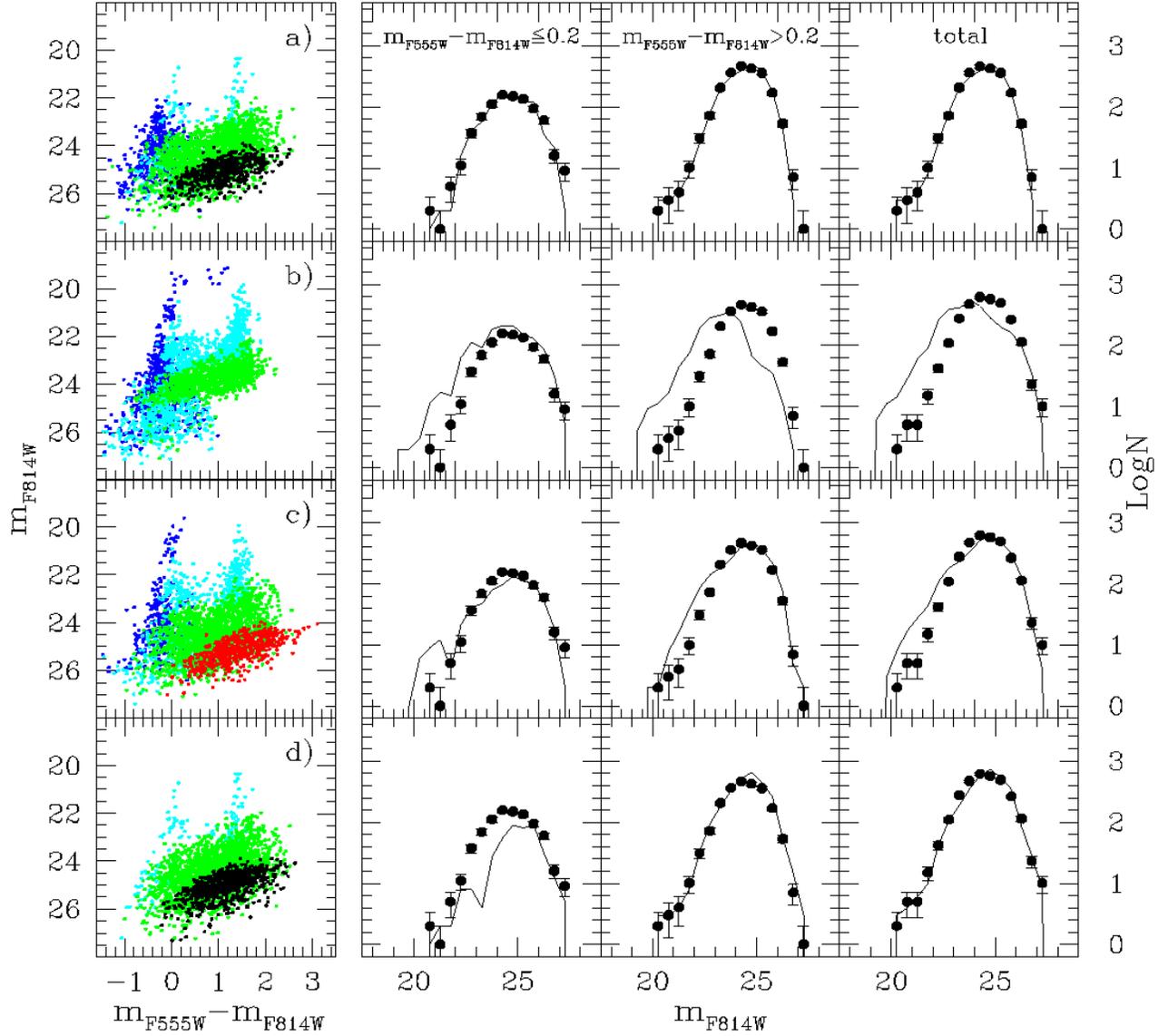}
\label{col_sim6}
\caption{Region 6: synthetic CMDs and LFs. 
 The stars have beeen plotted according to the following color coding:
 blue for t $< 10$ Myr, cyan for 10 Myr $\leq$ t $<$ 60 Myr,
 green for 60 Myr $\leq$ t $<$ 1 Gyr. Red and black denote the same age
 $>$ 1 Gyr but metallicities $Z=0.004$ and $Z=0.0004$ respectively.  	
 In panel a) we plot our best case (see text for details),
 characterized by: a young burst starting  2 Myr ago and still ongoing,
 a fluctuating and continuous SF from 1 Gyr to 10 Myr ago,
 a metal-poor 10 Gyr old population. 
 In panels b) and c) there are respectively the cases of two constant episodes
 starting  100 Myr and 5 Gyr ago and still active, both with $Z=0.004$.  
 Panel d) differs from the case of panel a) only for the lack
 of the young burst (2-0) Myr.
 All four cases assume $\alpha=2.35$.  
 Next to the synthetic CMDs we display the LFs. Dots are for the data, 
 continuous line for the simulations. From left to right, the LFs 
 refer respectively to \mvi $\leq 0.2$,  \mvi$>0.2$ and to the entire 
 color range.
}

\end{figure}

All the age-boxes appear populated, indicative of no obvious interruption of the SFR down to the look-back time. The look-back time itself is uncertain because the oldest stars in our CMD could be RGB stars whose age can range from $\sim$ 2 to 15 Gyr. Therefore we first explore the results of adopting a constant SFR, and treat the starting epoch as a free parameter. An example is shown in Fig.~\ref{col_sim6} panel b), which corresponds to a burst starting 100 Myr ago and still continuing. The adopted IMF slope is $\alpha$=2.35. Clearly, stars older than 100~Myr are present in our observed CMD. We find a starting epoch of at least 1-2 Gyr ago in order to account for the (faint) redder stars observed.  

In panel c) we show the case of a constant SFR starting 5 Gyr ago and still active.  The adopted IMF slope is $\alpha=2.35$. This simulation disagrees with the data by overproducing bright stars both in the blue and in the red  (see the LFs). Adopting a steeper IMF ($\alpha=2.6$), we improve the fit of the blue LFs at the bright end, but we also significantly underproduce blue stars in the range $22<$\mi$<25$. 

Since with a single episode we cannot successfully reproduce the observed feature, we applied multiple-episode cases with the usual age-box procedure.  We find that a constant SFR between 10 and 60 Myr ago does not provide a good fit to the LF of the stars in box B: we need to increase the SFR in the range 10-16 Myr with respect to that within 16-60 Myr. This conclusion results from the average of several runs and is unlikely to reflect stochastic effects. Besides, it agrees with the signature of the burst B1 in the same age range seen in Region 7. 

Intermediate ages (from $\sim$ 60 Myr up to $\sim$ 1 Gyr) are probed with the stellar counts within box C. These magnitude and color intervals include both helium burning stars younger than $\sim 100$ Myr and AGB stars with ages up to 1 Gyr. If we adopt a constant SFR over this whole period, the resulting synthetic CMD is underpopulated on the brightest blue loops. Hence, more power in the episode $\sim$ (100-60) Myr is needed.

The simulated number of stars with age between 10 Myr and 1 Gyr does not account for all the detected objects. We still need to further populate both the blue plume and the (faint) red portion of the CMD. The latter component can only be ascribed to a population older than 1 Gyr, i.e., RGB stars with some contribution from AGB stars. 

RGB stars are poor age indicators because their colors are only weakly dependent on age. In addition, the age-metallicity degeneracy introduces an additional uncertainty. Although the slope of the RGB depends mostly on metallicity (see Fig.~\ref{tracks}), this component in the CMD of Region 6 is completely embedded in the intermediate-age domain, so that its properties are difficult to single out. 

Having tested both the $Z=0.004$ and $Z=0.0004$ cases for different starting epochs, we prefer the less metal-rich and older (up to a Hubble time) model.

The partial models described so far are plotted in panel d) of Fig.~\ref{col_sim6} as cyan, green, and black dots for the age ranges 10 to 60 Myr, 60 Myr to 1 Gyr, and 1 to 10 Gyr ago, respectively.  We notice that the red portions of the CMD and the LFs are rather well reproduced, but a paucity of blue stars is apparent. We need a very young component populating box A, where the already modeled stellar generations account only for 14 \% of the observed stars. As already found in Region 7, we need a very young population to avoid an overproduction of the blue SGs. A burst starting 3 Myr ago and still continuing nicely accounts for the remaining objects. Again, the SFR in this episode turns out very high (see Tab.~\ref{tab_2}). The best case with all the derived SF episodes is shown in panel a) of Fig.~\ref{col_sim6}. 

We have explored the dependence of the fit on the IMF exponent. The synthetic CMDs that best reproduce the data are obtained with an IMF close to Salpeter or slightly steeper. For a instance, the CMD of panel a) of Fig.~\ref{col_sim6} has $\alpha=2.35$. A flattening at the high-mass end of the IMF tends to overproduce the number of bright MS stars. We can exclude an IMF as flat as $\alpha=1.5$.

We have also performed tests with quiescent intervals between the SF episodes. The intervals do not produce any difference on the resulting CMDs when added at old ages. However, we can exclude the presence of quiescent phases between the young episodes. If a quiescent period between 16 and 60 Myr ago is assumed, the synthetic CMD presents a gap in the red SGs distribution similar to that obtained in Region 7. This is inconsistent with the data.  

\begin{figure}
\figurenum{5}
\epsscale{0.40}
\plotone{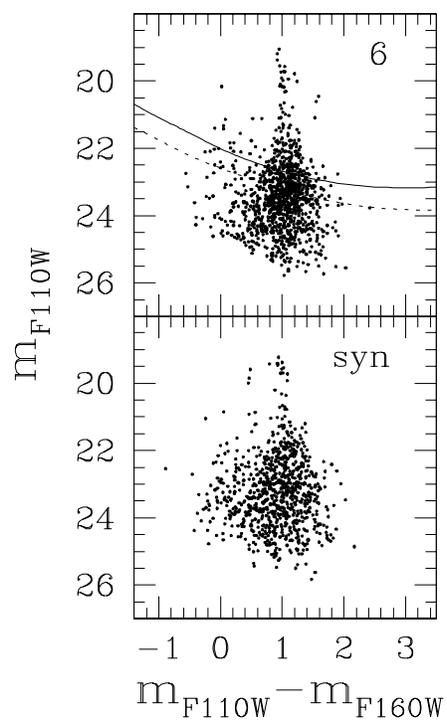}
\label{cm_jh_6}
\caption{In the top panel we show the observed NIR CMD for the portion
 of Region 6 covered by the NIC2 camera (only a small area is left out).
 The CMD contains 1079 objects (we have removed 8 objects corresponding 
 to the {\it pearl necklace}). 
 We overplot on the data the completeness levels at 75 
 \%  (solid line) and 50 \% (dotted line).
 In the bottom panel there is the synthetic
 CMD obtained adopting the SFH derived for Region 6 from the optical data.
 The synthetic NIR CMD reproduces only 78 \% of the observed objects.}
\end{figure}

The NIR CMD obtained with our best model (Fig.~\ref{col_sim6}, panel a)) is shown in Fig. ~\ref{cm_jh_6}. The overall morphologies of the empirical and synthetic CMDs appear quite similar, confirming the self-consistency of our results. However, we underproduce the objects detected on the NIC2 frames: only 78 \% of the 1078 objects in the empirical CMD are present in the synthetic CMD. The simulation lacks red stars at $22 \leq$ \mj $\leq 24.5$, and faint stars at all colors. This discrepancy can be ascribed to a combination of two effects. First, the simulator does not account for stars in the TP-AGB phase: if this component is better detected in the NIR than in the optical, the applied SFR underpredicts the synthetic stars in the bright AGB phase on the NIR CMD. Second, the NIC2 camera has larger pixels and a more patchy background. The artificial star experiments can yield inadequate photometric errors and incompleteness factors. In particular, multiple blending of stars is not considered, while it may well be at work in the NIC2 images. In other words, part of the objects detected at faint IR magnitudes could be multiple blends, which are not accounted for in the simulations. 

 \subsubsection{Region 5}

Fig. ~\ref{5obs} displays the observed \mi, \mvi \ for Region 5. The blue and the red plumes are still visible, but they are not so well defined as in Regions 6 and 7. The brightest star of the red plume is located at \mi $\sim 20.2$, while the blue plume appears much fainter, with its brightest object at \mi $\sim 22.1$. At \mi $>22.5$ we recognize blue loop stars, but this phase starts to be populated considerably only at \mi $\sim 24$. The dramatic drop of the crowding level with respect to the previous regions allows the detection of many faint, low-mass stars in the RGB phase. This phase has the highest population density in the empirical CMD. At \mi~$>$~24, we recognize a horizontal feature extending from the TRGB to the red, up to \mvi $\sim$ 3. Similar features have been found in other DIrr, like NGC~6822 (Gallart et al., 1996) and VII Zw 403 (Lynds et al., 1998; Schulte-Ladbeck et al., 1999a). They are thought to be caused by stars in the TP-AGB. Since we do not simulate this evolutionary phase, the determined SFR at intermediate ages is more uncertain.

Similar to the case of Region 6, the fact that all the age-boxes appear populated is indicative of no significant interruption of the SF down to our look-back time. The presence of RGB stars in the observed CMDs implies the presence of stars older than 1 Gyr.

\begin{figure}
\figurenum{6}
\epsscale{}
\plotone{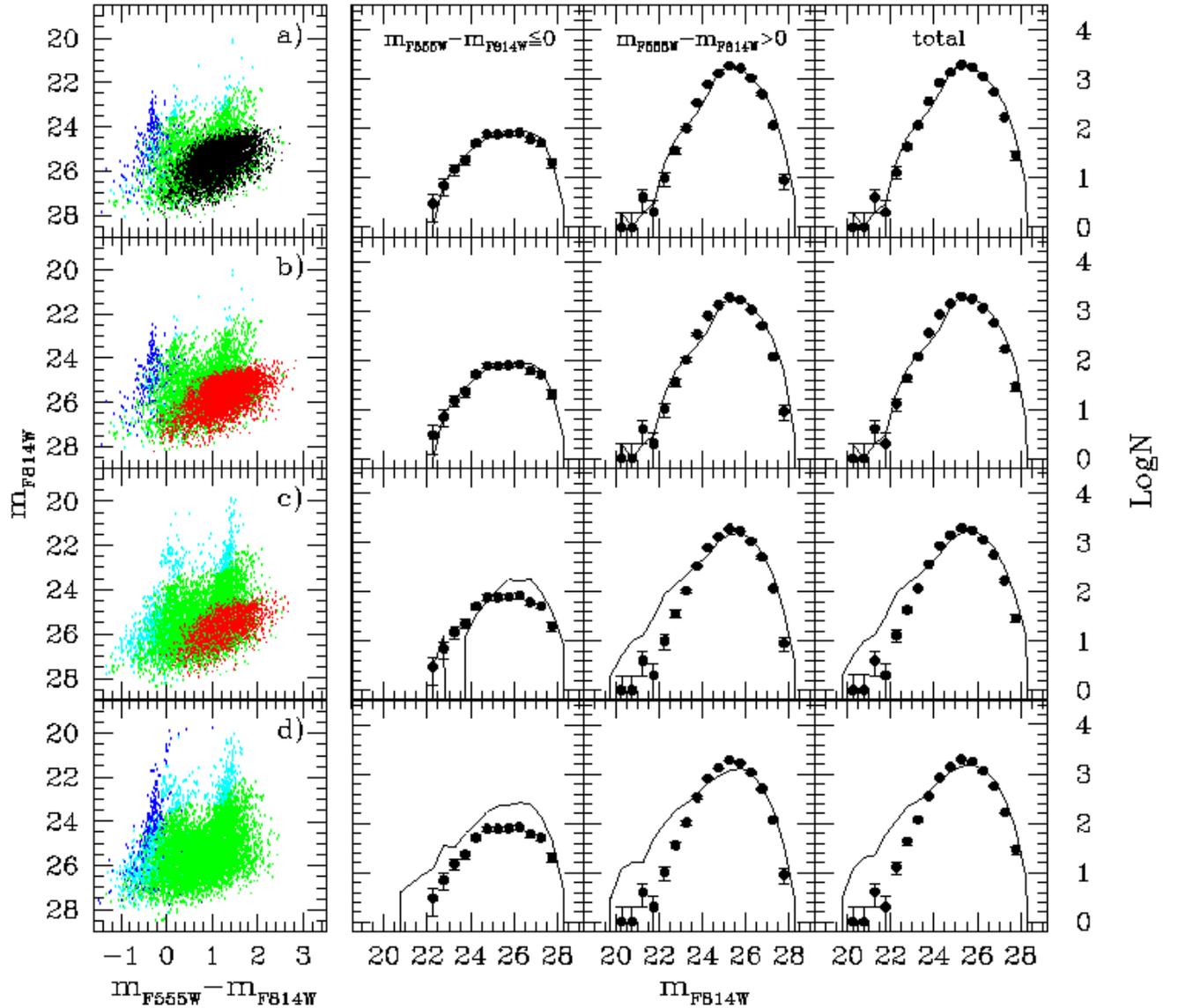}
\label{col_sim5}
\caption{Region 5: synthetic CMDs and LFs.
 The color coding is the same adopted for the previous regions. 
 Panel a) shows one of our best models, characterized 
 by: a young burst starting  2 Myr ago and still active,  
 a continuous SF from 1 Gyr to 50 Myr ago,   
 a metal-poor and 15 Gyr old population.
 The case in panel b) is similar to panel a) except for
 the old population, obtained with $Z=0.004$ and a 2 Gyr old start.
 In panel c) there is a constant episode from 2 Gyr to 10 Myr ago.
 In panel d) we have assumed a constant SF from 1 Gyr to now.
 All the cases assume $\alpha=2.35$.
 From left to right, the LFs refer respectively to \mvi $\leq 0$,  
 \mvi$>0$ (note the use of different cuts to better enhance the gap 
 in panel c) due to the lack of SF in the last 10 Myr)
 and to the whole color range.}
\end{figure}

\begin{figure}
\figurenum{7}
\epsscale{}
\plotone{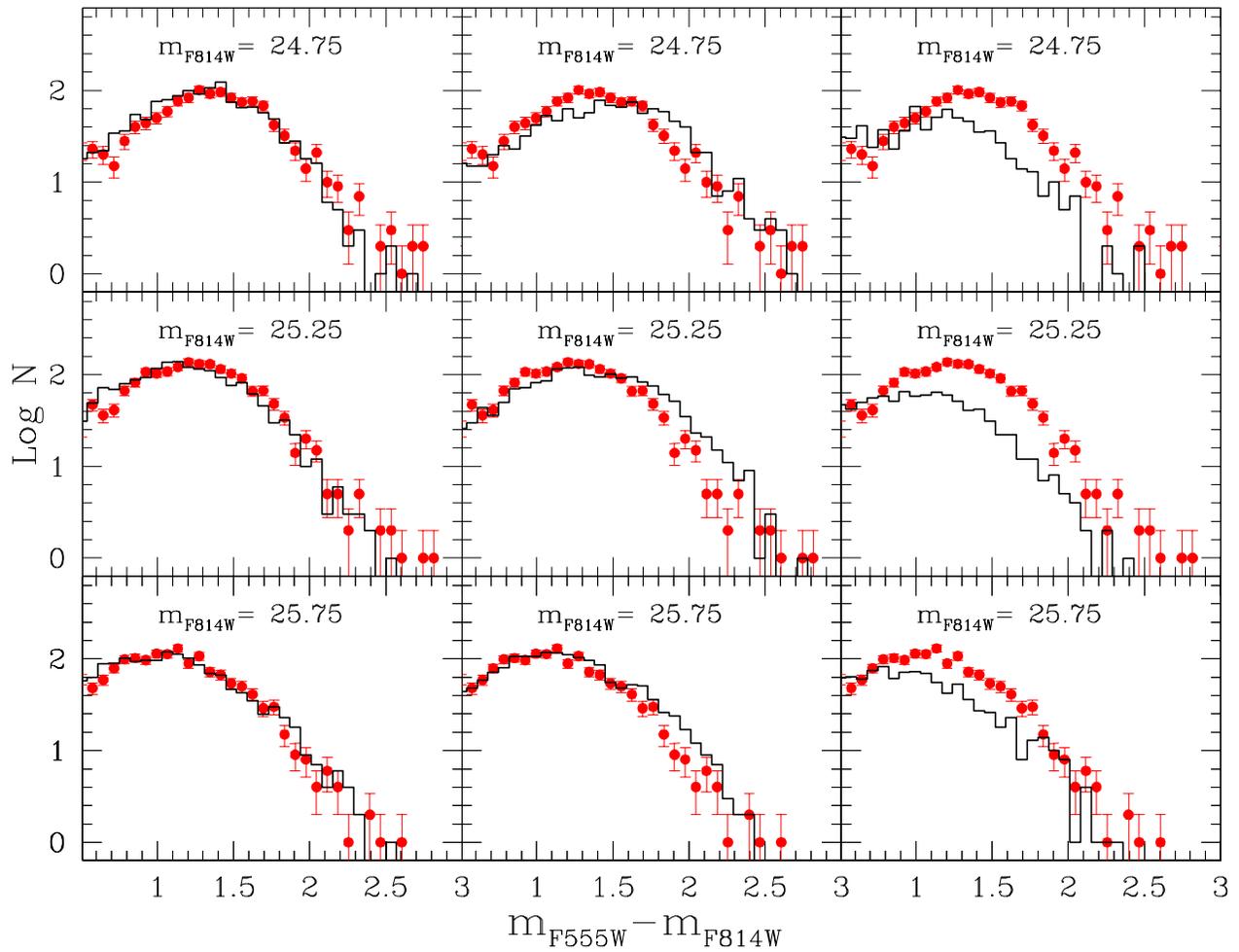}
\label{hist_5}
\caption{Region 5: color distributions for bins of 0.5 mag  
 for the red portion of the CMD within 24.5$<$ \mi$<$ 26.
 A label inside each panel gives the half bin value. 
 Red dots are for data, continuous black line for the simulations.  
 The left and central vertical sets of panels display the cases of 
 old-metal-poor and young-metal-rich RGBs (panels a) and b) of 
 Fig.~\ref{col_sim5}). 
 Right panels correspond to case d), where no stars older than 1 Gyr
 are present.}

\end{figure}

We plot in panel d) of Fig.~\ref{col_sim5} the case of a single episode starting 1 Gyr ago and still active. We required such an episode to account for the total number of observed stars. The lack of the RGB phase is evident in the model, particularly in the color distributions of Fig.~\ref{hist_5} (right column): red stars at magnitudes fainter than \mi $\sim$ 24.5 are significantly underproduced. In addition, we notice the overpopulation of the blue plume at all magnitudes, and the excess of bright red synthetic stars. A second example is shown in panel c) of Fig.~\ref{col_sim5}, where we plot the result of a constant SF starting 2 Gyr ago and terminated 10 Myr ago, again for $\alpha=2.35$. The major disagreement is the overproduction of both red SGs and faint blue stars, while the adopted termination 10 Myr ago causes the lack of blue MS stars at $23<$\mi$<24$. This suggests some younger SF within the past 10 Myr.

We have explored the results of constant SF scenarios for different starting/ending epochs without reaching a satisfactory agreement between the CMDs and their corresponding LFs. To achieve a fair representation of the data, we have then applied the age-box procedure. A single episode (see Tab.~\ref{tab_3}) in the interval (50-10) Myr, aimed at reproducing  the counts in the box B, fits well the observed distribution of the RSGs.

We use the counts in a bluer portion of box C (\mvi$<1.8$)
to constrain the partial model for the intermediate age population. The redder stars in this magnitude bin are assumed to be TP-AGB stars. This dividing line is rather arbitrary. Therefore we also constructed models with different partitions (see below). The sum of the two partial models (50-10 Myr in cyan $+$ 1000-50 Myr in green, panels a) and b) of Fig.~\ref{col_sim5}), amounts to 42 \% of the total number of objects in the observed CMD. Again, we need extra components, both in the blue and in the red. 

In the young age-box A the two partial models account for 12 \% of the observed stars. We thus simulate a young model to match the counts in this box. As found for Regions 7 and 6, the starting epoch of the young model cannot be older than 3~Myr, otherwise we obtain too many bright blue SGs. The stars plotted in blue in panels a) and b) originate from a burst starting 2 Myr ago and still continuing.

We investigated the metal poor ($Z=0.0004$) and metal rich ($Z=0.004$) scenarios with different starting epochs for the component older than 1 Gyr. Panel b) of Fig.~\ref{col_sim5} shows a total model in which the stars plotted in red have $Z=0.004$ and are generated in an episode from 2 to 1 Gyr ago. The corresponding color distributions are displayed in the central column of Fig.~\ref{hist_5}: since the RGB colors turn out too red (especially the peak color at the RGB Tip), a metal-rich and young burst has low probability.

On the other hand, our data are consistent with the opposite assumption of a metal-poor and old RGB. This case is investigated in panel a) of Fig.~\ref{col_sim5}, where the stars plotted in black belong to an episode starting 15 Gyr ago and terminating 1 Gyr ago. The color distributions of the relative total model (left column of Fig.~\ref{hist_5}) fit the data very well. The intermediate age component in the CMD (at \mvi $>1$ and magnitudes brighter than the TRGB) extends in a vertical feature. This is quite different from the distribution of the observed stars in the same region of the diagram. The discrepancy may result from inaccurate color transformations for such cool stars. It is worth mentioning that $\sim$ 35\% of the synthetic population in box C consists of E-AGB stars. In addition, the model is not meant to reproduce the TP-AGB population, which is also reflected in the paucity of synthetic objects at \mi $\sim$ 24 in the red LF.

Both the stellar distribution and the LF in this region of the CMD are sensitive to the SFR at intermediate ages. We have thus computed several models to improve the fit, and to establish a range of possible solutions. The results are summarized in Tab.~\ref{tab_3}. A constant SFR throughout the whole period is compatible with the data, as well as (small) fluctuations around this level. The upper limit to the SFR in the 0.5 to 1 Gyr age bin corresponds to a model required to reproduce the total number of objects in the $23<$\mi$<24$, 0.5 $<$ \mvi $<$ 3 box. This model, producing the highest number of objects in the region populated with RGB stars as well, corresponds to the lower limit to the SFR rate in the old age range indicated in Table 3.

As found for the previous regions, a Salpeter or slightly steeper IMF provides a good fit to the data, while flatter slopes, such as $\alpha=1.5$, overpredict the number of massive stars at the brightest bins.  In contrast, a steep IMF ($\alpha=2.6$) leads to acceptable models. The rates are summarized in Tab.~\ref{tab_3}.

\begin{figure}
\figurenum{8}
\epsscale{0.40}
\plotone{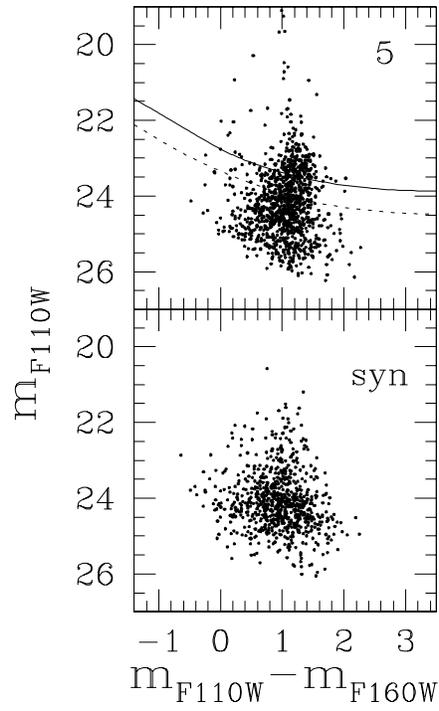}
\label{cm_jh_5}
\caption{The upper panel displays the NIR CMD for 
 the portion of Region 5 ($\sim 2/5$) covered by the NIC2.
 Two objects belonging to the spiral {\it pearl  necklace}
 have been removed, and we are left with 1137 objects.
 We overplot on the data the completeness levels at 75 
 \%  (solid line) and 50 \% (dotted line).
 In the bottom panel we plot the synthetic CMD obtained adopting the 
 SFH derived from the optical data. In this simulation,  
 only 70 \% of the observed objects are reproduced.}
\end{figure}

We have scaled the SFRs derived from the optical to the smaller area of the NIR field. This allows us to compare synthetic CMDs in the \mj, \ \mjh \ plane with the observations. Here we assume that the SFR per unit area is the same over the whole Region 5. It is possible that the episodes sampled in the optical CMD are not homogeneously spread over that region, in conflict with the above assumption. If, for example, most of the SF activity sampled in the optical occurred in the portion of Region 5 mapped with the NIC2 camera, we should adopt a relatively stronger SFR for our (\mj,\ \mjh) modeling. With this caveat, we show in Fig.~\ref{cm_jh_5} (bottom panel) a synthetic CMD obtained by scaling the rates of the optical CMD in panel a) of Fig.~\ref{col_sim5}. The overall appearance of the CMD seems reproduced, but the synthetic diagram is underpopulated with respect to the data, similar to the results in Region 6. For the same reasons as before, these inconsistencies are not unexpected.

\subsubsection{Regions 3--4 }

Regions 3 and 4 are treated together because of their similar populations and crowding conditions. The corresponding observed CMD in Fig.~\ref{5obs} shows a very prominent RGB feature, plus some scattered blue stars. The lack of a conspicuous blue component suggests a prominently old stellar population. We anticipate little sensitivity of the simulations to the IMF slope since the CMD mostly samples stars in post-MS phases. All the models presented here are for a Salpeter IMF.

\begin{figure}
\figurenum{9}
\epsscale{}
\plotone{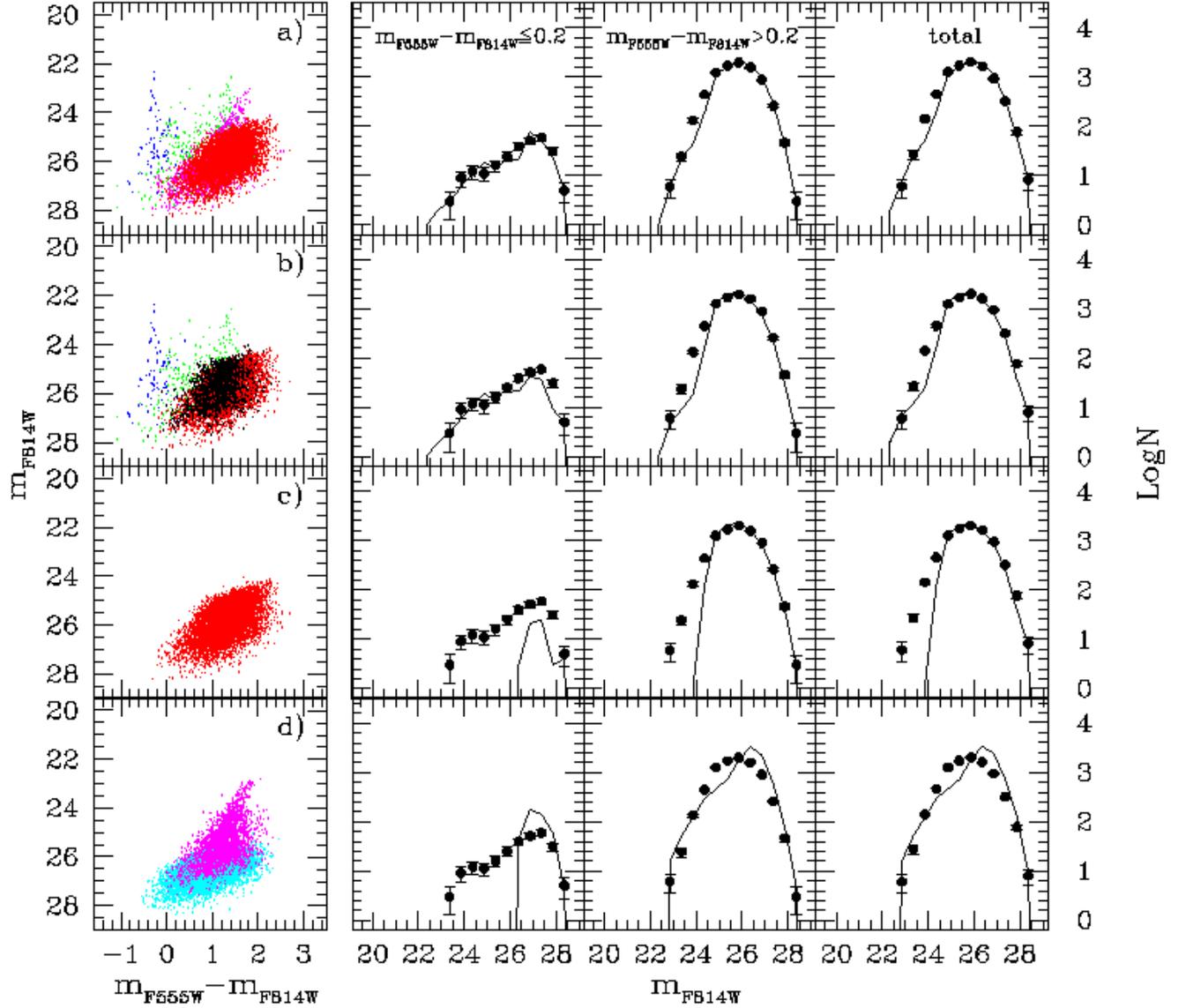}
\label{col_sim34}
\caption{Regions 3-4: synthetic CMDs and LFs.
 For panels a), b) and c) we adopted the following color coding:
 blue for t $< 10$ Myr, green for 10 Myr $\leq$ t $<$ 300 Myr,
 magenta for 300 Myr $\leq$ t $<$ 1 Gyr. Red and black denote the same age
 $>$ 1 Gyr but metallicities $Z=0.004$ and $Z=0.0004$ respectively.
 In panel a) both an intermediate-age component (1-0.3 Gyr)
 and an old metal rich component (3-1 Gyr)
 account for the RGB feature, with contributions of 0.25 \% and 
 0.75 \%. 
 In panel b)  
 the RGB feature consists of pure old stars  
 (1-15 Gyr), with the $Z=0.0004$ component accounting for 25 \% of it.
 Both case a) and b) assume a modest SF at ages younger than 300 Myr
 and a 2 Myr old burst.
 Panel c) is the case of a SF episode from 3 to 1 Gyr ago, based on
 $Z=0.004$ tracks.
 Panel d) displays a pure intermediate-age population (1-0.3 Gyr).
 The fainter cyan stars are core Helium-burning objects,
 the brighter magenta ones are their early AGB progeny.
 In all the simulations $\alpha=2.35$.
 Next to the synthetic CMDs we display the LFs. Dots are for the data, 
 continuous line for the simulations. From left to right, the LFs 
 refer respectively to \mvi $\leq 0.2$,  \mvi$>0.2$ and to the entire 
 color range.} 
\end{figure}

The (few) blue loop stars with \mi\ just fainter than 24 are consistent with a modest SF activity between 50 and 300 Myr ago. Older ages only provide red post-MS stars (see Fig.~\ref{tracks}). We need a very young burst of SF to generate the bluest stars. Partial models for ages less than 300 Myr are shown as green and blue dots in panels a) and b) of Fig.~\ref{col_sim34}. The poor statistics in this part of the CMD weakens the determination of the recent SF history in Region 3$-$4, so that the derived rates and epochs should be taken with care. Nevertheless, SF was definitely active in this region around $\sim$ 100 Myr ago, and continues to be active.

The RGB feature samples stars over a wide range of ages, including intermediate age stars. The latter component is better traced at magnitudes brighter than the TRGB, particularly in the ($24.5>$\mi$>23.5$,\mvi$> 1.2$) region, where we count $\simeq$ 430 objects. The partial model with ages between 50 and 300 Myr, constrained by the counts of blue loop stars, provides just 15 objects in this region. Thus, most of the objects in this box have to be ascribed to population older than 0.3 Gyr.

In principle, a significant fraction of the stars populating the RGB feature could belong to an intermediate age population. Fig.~\ref{col_sim34} panel d) shows that this is not the case. This simulation has been computed with an episode of star formation from 1 to 0.3 Gyr ago, forced to generate the total number of observed objects. The CMD shows a two-component population: the fainter (in cyan) consists of core-helium-burning objects, and the brighter (in magenta) of E-AGB stars. This distribution disagrees with the observed CMD. Moreover, the synthetic red LF peaks at too faint magnitudes. 

\begin{figure}
\figurenum{10}
\epsscale{}
\plotone{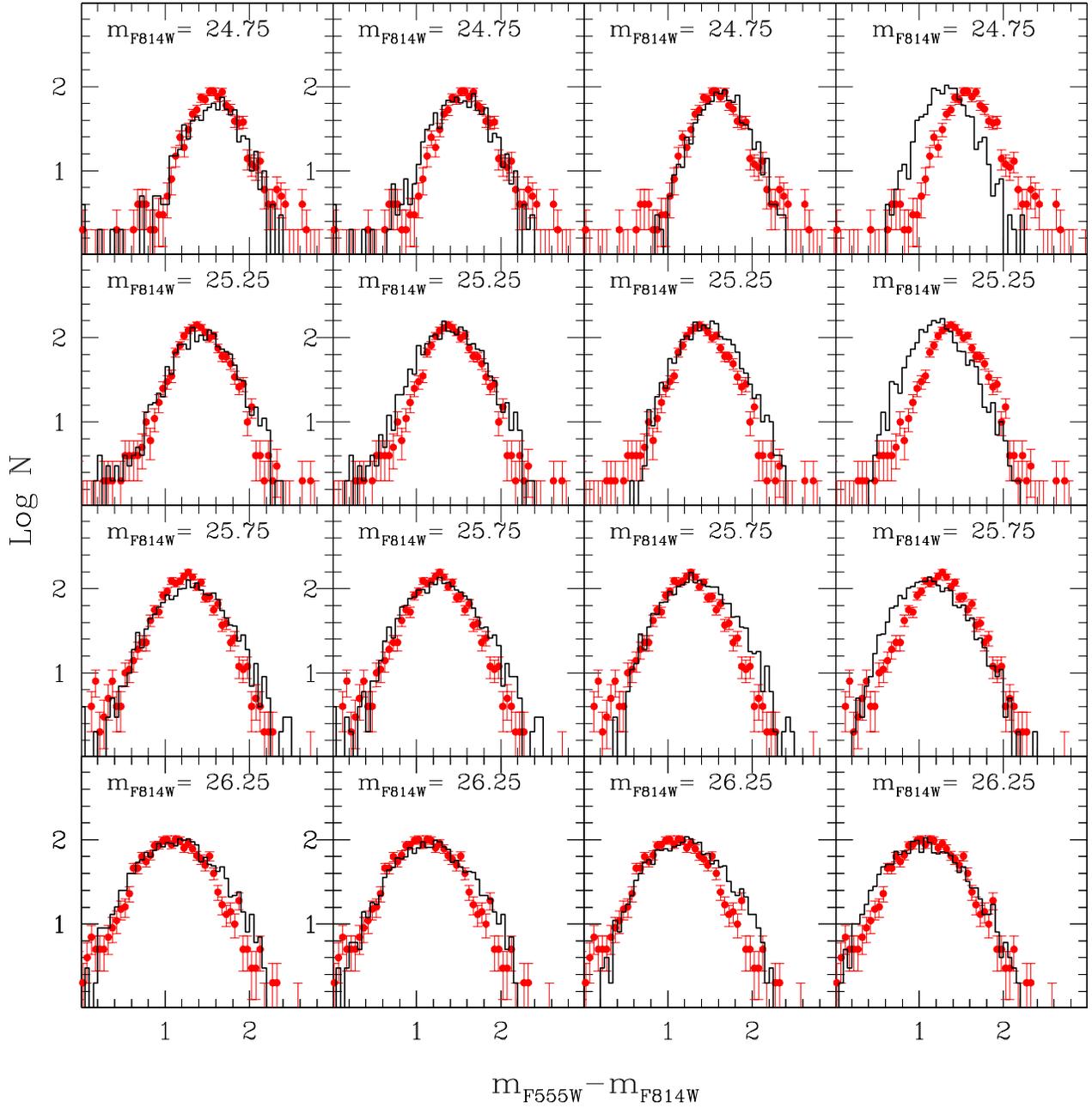}
\label{hist_34}
\caption{Regions 3-4. From left to right we plot in the
 vertical sets of panels
 the color distributions for cases a), b) and c) of 
  Fig.~\ref{col_sim34}. Right panels refer to a pure  
  $Z=0.0004$, 15 Gyr old population. 
  Red dots are for data, continuous black line for the synthetic cases.}
\end{figure}

 At the other extreme, we consider pure {\em old} population models. Panel c) in Fig.~\ref{col_sim34} shows the case of a SF episode from 3 to 1 Gyr ago, based on Z=0.004 tracks, and forced to provide all of the stars on the observational CMD. The corresponding color distributions in the upper RGB are plotted in the third panel (from the left) of Fig.~\ref{hist_34}. This model provides an acceptable description of the data, though some discrepancies are worth noting. There is a paucity of synthetic red stars brighter than the TRGB. This is at least partly caused by the lack of synthetic TP-AGB stars. Notice that blending of stars fainter than the TRGB is described in the simulations, and produces the synthetic stars brighter than \mi $\simeq$ 24.5. The color distributions show a modest excess of red stars fainter than 25. We consider this as a minor discrepancy because of the uncertainties affecting the color transformations for cool stars. However, extending this SF episode to older ages would lead to unacceptably red colors for the synthetic RGB.

Older ages could be consistent with the data if combined with lower $Z$. To test this option, we computed a model based on the $Z=0.0004$ tracks, starting 15 Gyr ago and terminating 1 Gyr ago. The corresponding RGB feature is too blue and too vertical on the CMD (see right panels of Fig.~\ref{hist_34}). We also tested models made of two components, one with $Z=0.004$ and ages between 1 and 3 Gyr, and the other with $Z=0.0004$ and ages between 3 and 15 Gyr, varying the relative fraction of stars in the two episodes (hereafter {\it Z-mixed} models). As expected, the RGB color distribution progressively shifts to the blue, as the ratio of the low-to-high $Z$ components increases. The effect, however, is not dramatic: for example panel b) of Fig.~\ref{col_sim34} shows one total model in which the population older than 1 Gyr corresponds to a {\it Z-mixed} model with 25 \% of the stars belonging to the $Z=0.0004$ component. The color distributions of the total model in panel b) are shown in the second column (from the left) in Fig.~\ref{hist_34}. We note that this model meets all the relevant constraints from the data, given the uncertainty of the color transformation of the cool stars and the lack of TP-AGB objects in the models. Considering that our {\it Z-mixed} models adopt an extremely low metallicity and a very wide range of ages, we conclude that the color distributions in this particular CMD could be fitted with a large variety of options for the age and metallicity spread. However, we find that the stellar mass in the RGB component of the CMD of Region 3$-$4 $is$ well determined. All the tested {\it Z-mixed} solutions yield a total mass in stars older than 1 Gyr in the range $(5.2-8) \times 10^7$ \MSUN\footnote{The quoted values for the mass in stars depend on the assumed IMF, i.e., a single Salpeter slope down to 0.1 \MSUN.}. The lower value applies to the case shown in panel c), the upper value to a pure $Z=0.0004$ model, with ages between 1 and 15 Gyrs. This is expected from basic stellar evolution arguments (see Greggio 2002): the production of evolved stars per unit mass of the parent stellar population decreases mildly with increasing age.

We also computed {\it age-mixed} models in which some stars in the RGB feature have an intermediate age. This can improve the fit in the magnitude range just above the TRGB. The adopted metallicity is $Z=0.004$ for both SF episodes, which are taken to occur between 0.3 and 1 Gyr, and between 1 and 3 Gyr. As the power in the younger of the two episodes increases, the red LF becomes less and less populated at the peak. A similar behavior occurs at the peak of the color distributions in the upper RGB. Both trends correspond to a less pronounced RGB Tip. Even in this case, though, the effect is not dramatic: panel a) in Fig.~\ref{col_sim34} shows a total model, in which 25\% of the stars populating the RGB have ages in the 0.3-1 Gyr interval. The corresponding color distributions are shown in the left panels of Fig.~\ref{hist_34}. We see that this model meets the major observational constraints. A larger proportion of stars in the 0.3-1 Gyr age range would worsen the fit around the TRGB. 

To summarize, we find several possibilities for the interpretation of the stars in the RGB feature: (i) a population with metallicity around $Z=0.004$ and an age range from 1 to a few Gyr; (ii) a composite population with a metallicity $and$ an age spread; (iii) a $Z=0.004$ population with a moderate intermediate age component. We tend to prefer this last model because it accounts for some of the stars detected above the TRGB. However, model uncertainties leave the intermediate age SF in this region quite unconstrained, as summarized in Table~\ref{tab_4}. The model shown in panel a) has a mass in stars older than 0.3 Gyr of $4.5\times10^{7}$ \MSUN: again, the mass in the RGB feature turns out to be a robust result. In contrast, the epochs at which this mass was converted into stars are relatively uncertain, and so is the corresponding SF rate.

\subsubsection{Regions 0--1--2}

For the same reasons as for Regions 3 and 4, we have treated together Regions 2, 1 and 0. The corresponding CMD is shown in the bottom panel of Fig.~\ref{5obs}. It is quite similar to that of Region 3$-$4, with an even larger fraction of stars in the RGB feature \footnote{The RGB of Region 0$-$1$-$2 is much steeper than in the other regions because of the smaller magnitude range of the plot.}. This suggests that the SF at intermediate and young epochs was very low. However, the bluest stars have colors \mvi $<0$, and some recent SF occurred even in these peripheral regions.

\begin{figure}
\figurenum{11}
\epsscale{}
\plotone{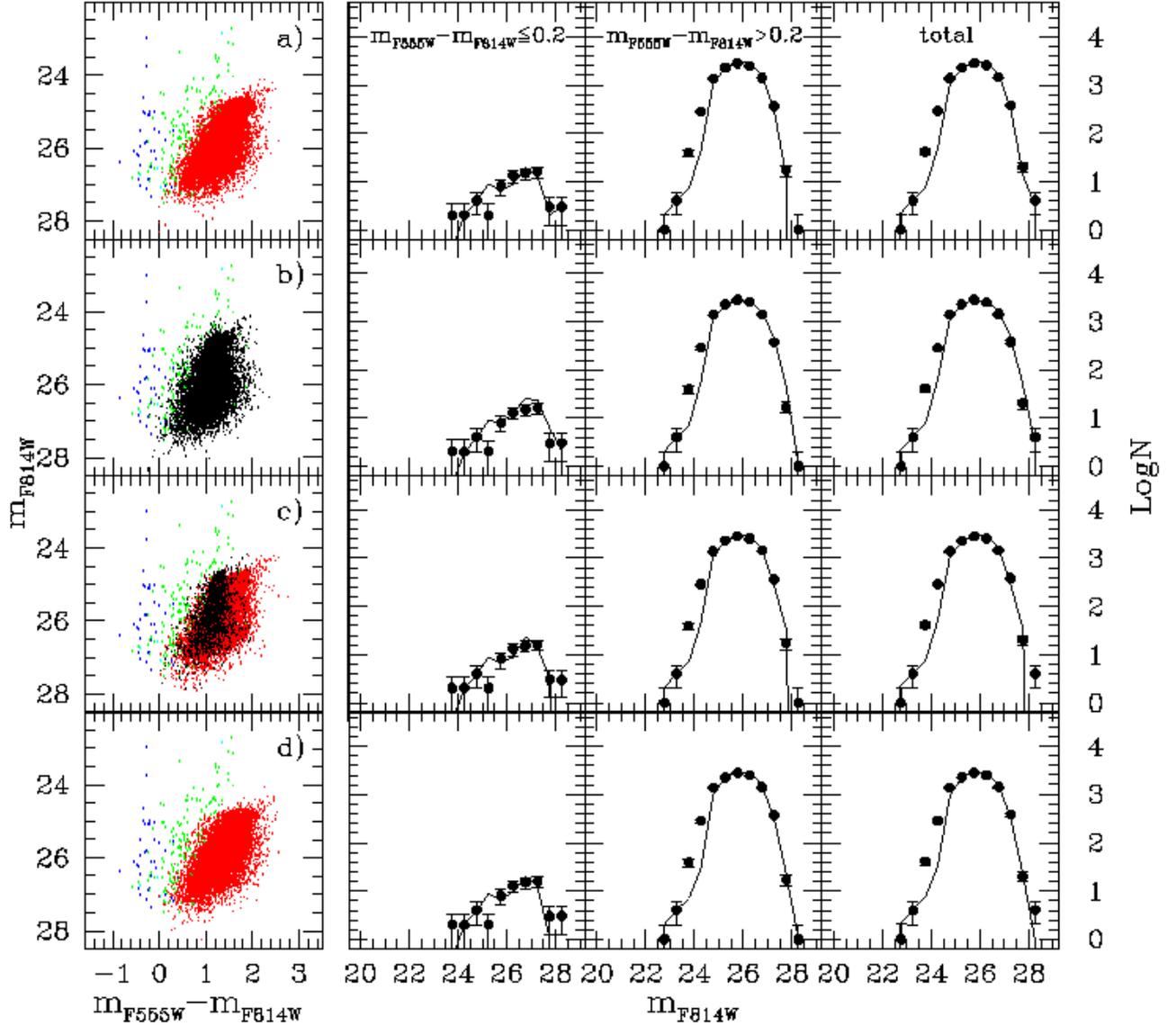}
\label{col_sim012}
\caption{Regions 0-1-2: synthetic CMDs and LFs.
 For the adopted color coding, see the previous regions.
 Our best model, plotted in panel a), assumes  
 a 5 Gyr old RGB with $Z=0.004$.
 In panel b) there is the case of a metal poor ($Z=0.0004$) 15 Gyr
 old RGB.
 In panel c) a small fraction (1/6) of the RGB stars has 
 $Z=0.0004$ and corresponds to the oldest ages, 
  the remaining 5/6 have $Z=0.004$. In panel d) we put a 4 Gyr
 gap between two episodes (7-6) Gyr and (2-1) Gyr; $Z=0.004$. 
 All four cases assume also a weak SF at ages $<$ 1 Gyr.
  $\alpha=2.35$.
  Next to the synthetic CMDs we display the LFs. Dots are for the data, 
 continuous line for the simulations. From left to right, the LFs 
 refer respectively to \mvi $\leq 0.2$,  \mvi$>0.2$ and to the entire 
 color range.}
\end{figure}

Our main goal is to derive the age and metallicity parameters of the old population for Region 0$-$1$-$2. The intermediate and young SFH cannot be confidently inferred due to the small number of objects detected in this age ranges. Nevertheless, it is evident that a SF activity has occurred at intermediate and recent epochs. Hence, we have constructed a few models for this component. An example of a good model is shown in Fig.~\ref{col_sim012} as blue and green dots, and the relative SF rates are given in Table~\ref{tab_5}. The blue dots belong to the current SF episode. The distribution of the observed stars bluer than \mvi =0 is still consistent with a very young burst starting about 3 Myr ago, similarly to the inner regions, but can also be reproduced by weaker continuous activity over the last 20 Myr. The green population accounts for the blue loop stars and part of the red stars brighter than the TRGB. It was generated in an episode active between 50 Myr and 1 Gyr ago. The relative SFR are quoted in Table~\ref{tab_5} and should be regarded as approximate. 

In comparison with the more central regions, the RGB feature in Region 0$-$1$-$2 is characterized by a smaller color spread. This is partly due to the very low crowding level. The superior quality of the photometry in this outer region provides better constraints on the metallicity distribution.

\begin{figure}
\figurenum{12}
\epsscale{}
\plotone{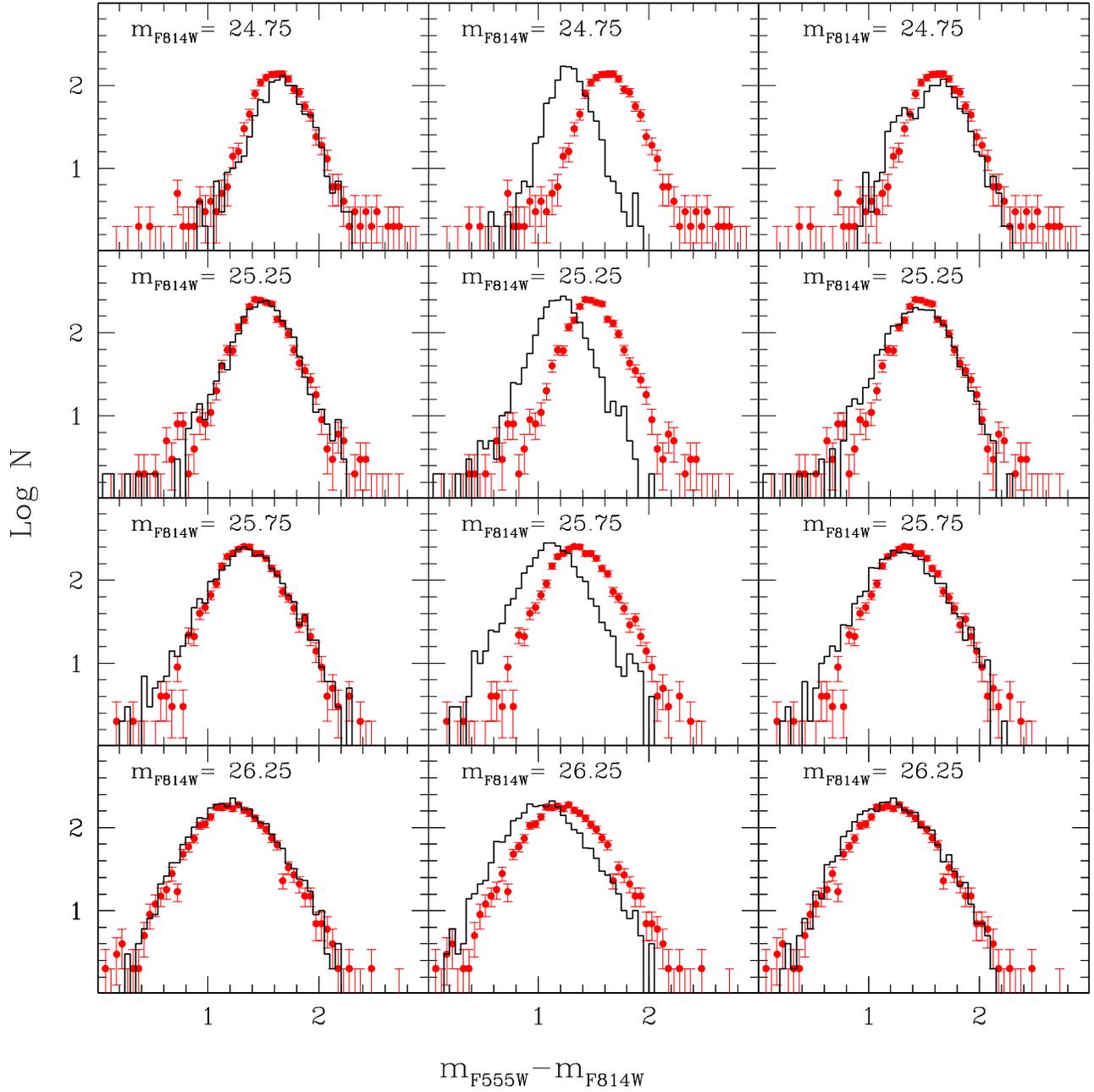}
\label{hist_012}
\caption{Regions 0-1-2: left, central and right vertical sets
 of panels refer respectively to a metal-rich 5 Gyr old case,
 a metal-poor 15 Gyr old case, a Z-mixed metallicity case
 (respectively panels a), b) and c) of Fig.~\ref{col_sim012}).
 Red dots are for data, continuous black line for the models. 
}
\end{figure}

In Fig.~\ref{col_sim012} we show representative cases of synthetic CMDs for a Salpeter IMF. In panel b), the simulated RGB (black dots) results from a single episode starting 15 Gyr ago and terminating 1 Gyr ago; the adopted metallicity is $Z=0.0004$. Even with such an old age, the very metal-poor evolutionary tracks produce an RGB that is definitely too blue and steep. This is evident in the color distributions displayed in the second column of Fig.~\ref{hist_012}. In panel c) we show a {\it Z-mixed} model: 1/6 of the stars have $Z=0.0004$ and ages 10 to 5 Gyr; the remaining 5/6 have $Z=0.004$ and ages 5 to 1 Gyr.

The immediate effect of using two different metalliciites is the presence of two separated RGBs in the CMD, whereas such a dichotomy is not observed (see also the right column of Fig.~\ref{hist_012} for the color distributions of the two RGBs). It is clear that the very metal-poor tracks by themselves cannot reproduce the observed RGB, as they did for the previous regions. Furthermore, even a relatively small fraction of stars in the very low metallicity component is excluded. On the other hand, the possibility of a stellar component with an intermediate metallicity between 0.004 and 0.0004 still exists.

In panel a) of Fig.~\ref{col_sim012} we show our best case, with the corresponding RGB color distributions in the first column of Fig.~\ref{hist_012}. Only the $Z=0.004$ tracks have been used (red dots), and an age range of 1 to 5 Gyr has been adopted. Such an epoch of SF turns out to be consistent with the observed RGB colors, while older ages tend to produce too red RGBs. We report the rates derived from our best model in Tab.~\ref{tab_5}. The total masses in stars older than 1 Gyr of the models shown in panels a) and b) are 6.8 and 8.3 $\times 10^{7}$ \MSUN, respectively. Again, we emphasize the robustness of this prediction.  

Finally, we tested the effect of quiescent intervals between the SF episodes in this region, whose photometry is of exceptional quality. In panel d) of Fig.~\ref{col_sim012}, the synthetic RGB is obtained from two separate episodes, both with $Z=0.004$: an older one from 7 to 6 Gyr ago, and a younger one from 2 to 1 Gyr ago. In spite of the very short duration of the two episodes compared to the 4 Gyr gap, no visible effect on the CMD is produced. The SFR derived at ages older than 1 Gyr should be considered as average values over a long period.

\section{The SF history of NGC~1705}

\subsection{Overview}

Tables ~\ref{tab_1} through ~\ref{tab_5}  list the derived SFH in the 5 zones of NGC~1705. 
Regions 7, 6, 5, 3-4, 0-1-2 are approximately located within
0.07, 0.2, 0.4, 0.8 and 3 kpc from the center of the galaxy.
The levels of the SFR are calculated with a mass range 120-0.6 \MSUN\ and extrapolated down to the lower mass limit of 0.1 \MSUN\ with a single slope. For the three innermost regions, where the simulations are sensitive to the IMF slope, we show the SFRs derived for those values of $\alpha$ which were found to agree with the data. 

For each SF episode we quote the starting and ending epochs and their 
uncertainties, and the range of SFR which results from our fit. 
For the most recent episodes, the range corresponds to variations in 
the starting and ending epochs, with the lower (upper) limit corresponding 
to the longer (shorter) duration. In the intermediate age regime, the 
range quoted for the SFR mostly reflects the options for the fit of the 
red stars at 23 $\lsim$ \mvi $\lsim$ 24.5. A fraction of these stars 
could be un-modeled TP-AGB objects. Solutions which maximize the number 
of synthetic stars in this region of the CMD correspond to the maximum 
production of intermediate-age stars. 
In turn, they also correspond to the minimum production of (older than 1 Gyr)
RGB stars, in order to match the total counts of red stars fainter than 
\mi$=$24.5.
As for the SFH at epochs older than 1 Gyr, the tracing stars cluster in the 
same region of the CMD, almost irrespectively of their age.
Therefore, we can only derive the average SFR over a long time interval whose boundaries are determined from the RGB star colors. Two sources of uncertainty affect the determination of the {\em average} SFR at old epochs: the quality of the fit to the intermediate-age component and the age-metallicity degeneracy. For the component older than 1 Gyr, we report the range in {\em average} SFR for all possible solutions. 

\begin{figure}
\figurenum{13}
\epsscale{}
\plotone{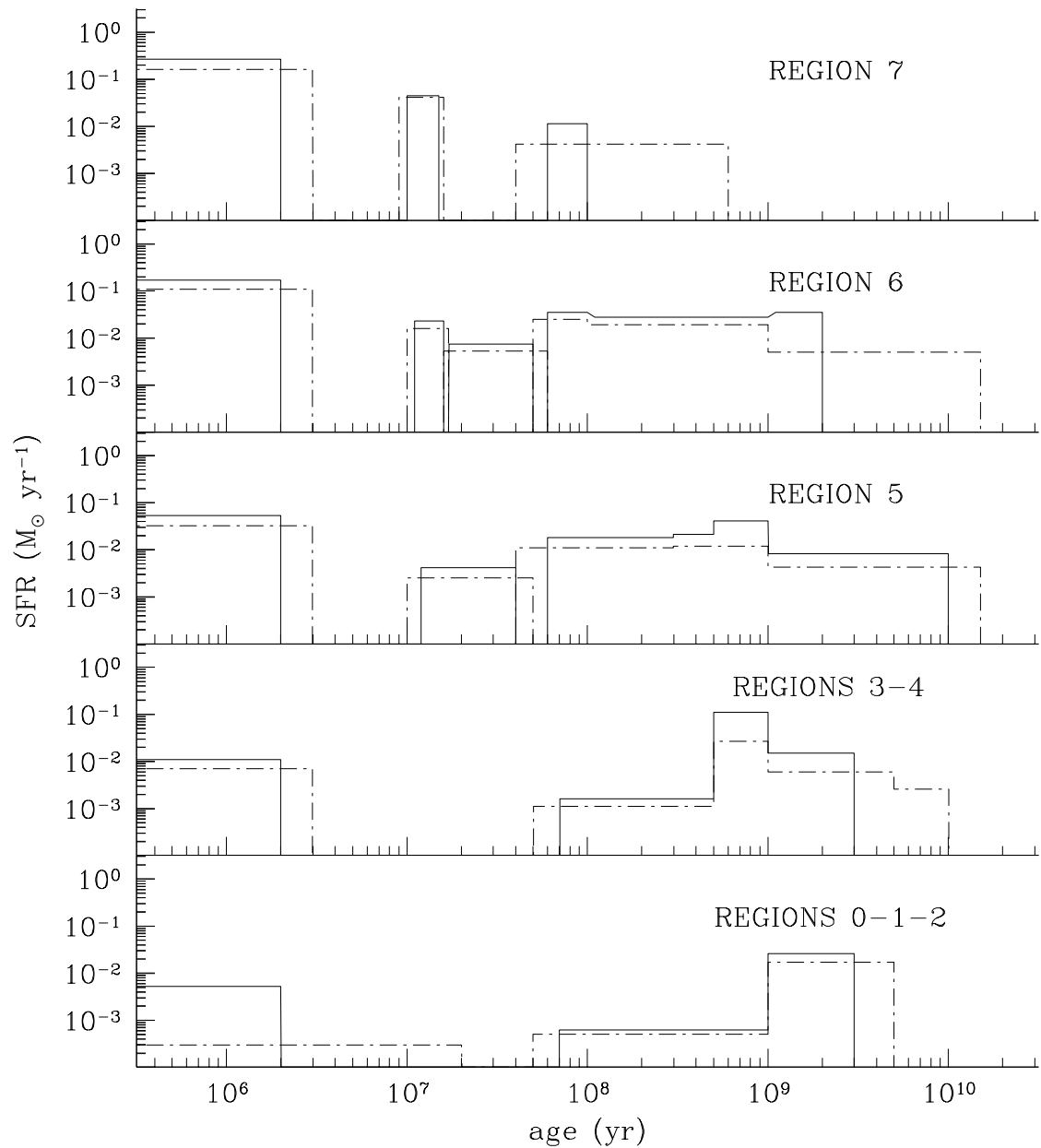}
\label{sfr_tot}
\caption{Star formation rate versus age (in logarithmic scale) for the 
 various regions of NGC~1705. The solid and dotted lines denote 
 respectively our upper
 and lower limits. The rates have been derived assuming 
 a Salpeter's slope from 120 to 0.1 \MSUN.}
\end{figure}

The results of Tables 1 to 5 are plotted in Fig.~\ref{sfr_tot}. The SFH is not the same over the whole galaxy: different zones turned out to experience different SF episodes, in terms of both the epoch and the strength of the SF activity. The young population dominates in the central regions, where we derive recent and relatively intense bursts from the CMDs. The strength of the youngest activity gradually decreases toward the periphery of the galaxy. The old population is spread over the entire galaxy, though the detection of the old stars is more difficult (and not always possible, as in Region 7) in the central regions. This is both because of the higher crowding level, and because of the larger component of young and intermediate-age stars in the innermost regions, which causes some confusion in the derivation of the earliest episodes of SF.  In the following sub-sections we describe our results for the recent, intermediate, and old SF. 

\subsection{Recent star formation}

A burst of SF starting $\sim$ 15 Myr ago and terminating $\sim$ 10 Myr ago (B1) is clearly detected in the field stars at the center of NGC~1705. The age corresponds to that inferred for the central SSC from previous photometric and spectroscopic studies (MFDC; Heckman \& Leitherer, 1997). The coincidence of the two SF events suggests a relation between the SSC formation and the field SF: SF may first have occurred in the SSC and then propagated to the adjacent regions through shocks induced by supernovae and stellar winds. B1 is very well traced in Region 7, whose CMD shows a definite upper boundary to the age of the stars generated in this episode. In Region 6, B1 is visible as an enhancement of the SFR over a more continuous and more moderate activity, starting $\sim$ 60 Myr ago. This continuous activity is visible also in Region 5, which does not seem to be affected by B1. Quiescent periods of the SF in the 60$-$10 Myr age range would appear as gaps in the distribution of the RGs.

 Most intriguing is the ubiquitous detection of the current burst B2, albeit with a rate decreasing from the center towards the outskirts. This burst is well separated from B1: all but one region do not show stars with ages between $\sim$3 and 10 Myr which would appear as blue SGs. Only in the outermost of our analyzed regions are the bluest stars compatible with a large range in age. This, however, reflects the poorer diagnostics from the small number of blue objects sampled. The data are consistent with a quiescent period of several Myr, followed by a new burst of SF with a radially decreasing strength over most of the galaxy.

The preferred IMF slopes are in the range 2.35 - 2.6, except for Region~7, where the best agreement is found for $\alpha$ between 2 and 2.35. With a slope steeper than Salpeter, the intensity of the two young bursts becomes slightly higher, as the IMF predicts a relatively larger number of low-mass stars and the rates for the recent episodes are constrained by the high mass stars on the CMD. Vice versa, with flatter IMFs, we derive smaller SFRs for the young bursts. For each episode, the differences between the SFRs corresponding to the possible values of $\alpha$ (see Tab.~\ref{tab_1} and following) should be taken as our uncertainty on the derived rates. 

 \subsection{Star formation at intermediate and old ages}

For epochs between 50 Myr and 1 Gyr ago, we derive an almost continuous and fluctuating SF, with modest variations in the rates. We exclude the presence of big gaps between the SF episodes at recent times. 

 In the age range 0.3$-$1 Gyr, the determination of the average SF rate is complicated by the difficulty of describing the TP-AGB phase. Though short lived, this evolutionary stage is well sampled in some of our observational CMDs, especially in Region 5 with the nearly horizontal feature stretching above the TRGB. The SFR levels in the Tables ~\ref{tab_1}--~\ref{tab_5} and Fig~\ref{sfr_tot} include this uncertainty. The intermediate age population is mostly present in the three central regions. Moving out of Region 5, the intensity of the SF episodes younger than 1 Gyr gradually decreases, and Regions 0$-$1$-$2 experienced only a very weak activity in the interval (1000-50) Myr.

We have low time resolution for ages older than 1 Gyr because populations with a wide range of ages occupy the same regions of the CMD and suffer from the age-metallicity degeneracy. Nevertheless, we derive rough constraints on both the age and the metallicity of the old population. The high crowding level of Region 7 prevents us from detecting stars in the RGB phase, and its CMD does not give any information about the early activity. The incompleteness level drops sufficiently in Region 6 to allow the detection of some RGB stars. Here, the very low metallicity tracks ($Z=0.0004$) reproduce the RGB colors better, in combination with a very early start of the activity. The high metallicity solution in our set of tracks provides too many faint red stars. However, this conclusion is weakened by the large fraction of intermediate-age stars contaminating the same CMD regions where we detect RGB stars. In Region 5, the RGB is better defined, but still with large photometric errors and contamination by younger stars. We favor the low metallicity solution with an early start for the SF activity (15-10 Gyr) with more confidence than in Region 6, since the peak of the color distribution of the bright RGB stars is very well matched with the low Z tracks. In Regions 3 and 4, the contamination of the RGB by younger stars drops significantly. There, the low Z (0.0004) models are too blue, and we find a good fit to the data using the high Z (0.004) models in combination with relatively late starting epochs. We tested {\it Z-mixed} models, and found that the data are well described by a model in which 25\% of the RGB stars are old and very metal-poor and 75 \% are metal-rich and younger. Larger proportions of very metal-poor stars are not supported, as they imply bluer color distributions for the RGB stars.

The three outermost regions (2,1 and 0) have a very low crowding level and almost no contamination by young and intermediate-age stars. The RGB is much narrower than in the other regions and well defined. A single metallicity population with $Z$=0.004 and ages between 1 and 5 Gyr yields a very good fit to the data. The {\it Z-mixed} models show that only a small fraction of RGB stars may belong to an older and very metal-poor population. 

We acknowledge that our investigation on the metallicity of the RGB stars is limited by the use of just two extreme values of $Z$. A variety of solutions are obviously possible, ranging from a single metallicity population with an age spread, to a wide metallicity distribution coupled with a convenient age-metallicity relation. However we find clear evidence for a metallicity gradient, with $Z$ increasing outward. This conclusion rests upon the very good fit obtained with the Z=0.004 tracks in Regions 3$-$4 and 0$-$1$-$2, and on the fact that this same set of tracks provides too many red stars fainter than the TRGB in Regions 5 and 6. Our solution is mostly driven by the RGB color, which becomes progressively redder going outward. The large metallicity corresponds to a relatively young age for the RGB stars in the outermost regions. SF at very early epochs is possible, in combination with lower metallicities, but a conspicuous proportion of stars as old as the Hubble time is ruled out.

\subsection {Summary}

In Table \ref{tabsummary} we summarize the average SF rates in representative 
age bins for five different regions. 
From 7 to 0 the regions are located at increasing distances from the 
center. This allows us to map the SFH occurred in NGC~1705.
The average rates do not suffer from the uncertainty of the starting and 
ending epochs of the SF episodes, and are better suited for characterizing 
the SF history in space and time.

The two most recent bursts of SF are very concentrated. The rate decreases by one order of magnitude from Region 6 to Region 5. The burst B2 is much stronger than B1; the total SFR in B2 is $\sim$ 0.3 \MSUN/yr for a Salpeter IMF, which is quite a high value. As discussed in Sect.~3, a lower rate would require longer time scales in order to reproduce the observed number of blue stars. This is highly unlikely, as it would produce bright blue {\em supergiants}, which are not observed.

At epochs between 15 to 50 Myr ago the activity was rather low, and we can trace the SF only in Regions 6 and 5. In contrast, we find stars with ages in the range 50 Myr to 1 Gyr throughout the whole galaxy. The SFR at intermediate ages is stronger in the central part of the galaxy. The determination of the rate at these epochs in Region 7 is substantially affected by crowding, and should be regarded as a lower limit.

RGB stars with ages above 1~Gyr are found at all galactic radii where the crowding conditions allow their detection. The SFR averaged over the fixed age range of 1$-$5 Gyr is very similar in all the galactic regions. We caution that the poor time resolution for the RGB feature limits constraints on the epochs related to this feature. Nevertheless, the mass in stars older than 1 Gyr is approximately the same in the various regions investigated. Since the area of the inner regions is small, this suggests that at old epochs the SFR per unit area was centrally concentrated.

In Table \ref{tabsummary} we also list the total SFR in the different age bins. We use these estimates to infer the mass that went into stars at the various epochs. The total mass in the current burst is $\sim$ 10$^{6}$\MSUN, comparable to that of a rich globular cluster. The total SFR in B1 is $\sim$ 7 $\times 10^{-2}$ \MSUN/yr, corresponding to an astrated mass of $\sim$ 3.5 $\times 10^{5}$ \MSUN. This is significantly larger than the total mass in the SSC derived by Ho \& Filippenko (1996). Therefore, a fair fraction of the interstellar gas was used to form field stars during the SF episode, which generated the SSC.

The mass that went into stars between 50 Myr$-$ 1 Gyr ago totals $\sim 6 \times 10^{7}$. Most of the astrated mass is in the component older than 1 Gyr, for which we give a lower limit of 2.2 $\times 10^{8}$. The astrated mass is larger than the current mass because it does not take into account the stellar mass returned to the interstellar medium. The correction factor however is not large: for Salpeter IMF, after 3 Gyr, $\approx$ 30 \% of the astrated mass went back to the ISM.

\section{Discussion and conclusions}

 We have inferred the SFH of NGC~1705 for the five different zones of the galaxy described in T01: Region 7 (within 0.07 kpc from the center), Region 6
(0.2 kpc), Region 5 (0.4 kpc), Regions 3$-$4 (0.8 kpc)
and Regions 0$-$1$-$2 (up to 3 kpc from the center). The division is driven by the different crowding conditions and stellar populations of the  galactic fields. Obviously, studying the SFH in different locations is meritorious for understanding the SF process itself (e.g., if at early epochs the activity started in particular regions, what triggers the SF, etc.). 

Our first major result concerns the age of NGC~1705. Past photometric studies (MDFC; Quillen et al. 1995) predicted the presence of an old population, besides the young one, from the red colors of an external region with low surface brightness. Our simulations confirm that NGC~1705 is not a young galaxy, but started forming stars several Gyr ago. The same result applies to all the other dwarfs resolved into single stars and studied so far (e.g., VII Zw 403, Schulte-Ladbeck et al. 1999b, Crone et al. 2002; Mrk 178, Schulte-Ladbeck et al. 2000; see also the reviews of Grebel 1999 and Tosi 2001). Even in IZw18, the most metal-poor BCD ever observed, there is evidence for stars born at least several hundreds of Myr ago (Aloisi et al. 1999; \"Ostlin 2000; Hunt et al. 2003).  Our data are not deep enough to reach the main-sequence turn-off (TO) or the horizontal branch (HB). Therefore we cannot obtain unambiguous results on the age of the old population. The uncertainty on the old components grows toward the most central regions, where the high crowding level and the significant presence of intermediate-age and young stars preclude the detection of a well-defined RGB. Despite the age-metallicity degeneracy, we tried to disentangle these two parameters by computing simulations with the $Z=0.004$ and $Z=0.0004$ Padova tracks. Simulations at intermediate metallicities would not be worthwhile, given the effects of incompleteness and photometric errors. 

A 5 Gyr old SF is consistent with all the data. We find evidence for a metallicity gradient of the old population from the center to the periphery of the galaxy. In the outermost regions, the RGB star distributions are compatible with a single metallicity of $Z=0.004$. Towards the center, the best models are obtained with the lower Z tracks in combination with older RGB stars. This result may have important implications for the SF process at early times: the activity may have first started in the center of the galaxy, followed later by SF in the external regions in a more metal rich environment. If so, the RGB population should be older in the center than in the periphery, but it is not possible to test this hypothesis from our data. Evidence for a similar metallicity evolution on the RGB was found also in UGCA 290 (Crone et al. 2002), another BCD studied with the method of the synthetic CMDs.

The SF has been almost continuous from 1 Gyr ago until now, with modest fluctuations of its level. We clearly detect a gap in the SF activity between 3 and 10 Myr ago over the whole galaxy. Another gap, from $\sim$ 15 to $\sim$ 40 Myr is evident only in Region 7. Such short gaps are easily hidden in the CMD at ages beyond 100 Myr. On the other hand, there is no evidence of quiescent periods of $\sim$ 0.1 Gyr duration over the last 1 Gyr. We denote the recent and intermediate age activity as a {\it gasping} SF, as if the galaxy once in a while has an {\it apnea}.
The absence of long quiescent intervals between   the SF episodes is a result common to all the other BCDs resolved into single stars: the traditional picture of intense short bursts separated by long quiescent phases does not seem to apply to the recent life of these objects (Marlowe et al. 1999). However, quiescent periods as long as few Gyr could be present at epochs older than 1 Gyr. Even in Regions 0$-$1$-$2, where we have the lowest photometric error, a 4 Gyr interruption at early epochs is equally consistent with our RGB data as a continuous SF.

\begin{figure}
\figurenum{14}
\epsscale{}
\plotone{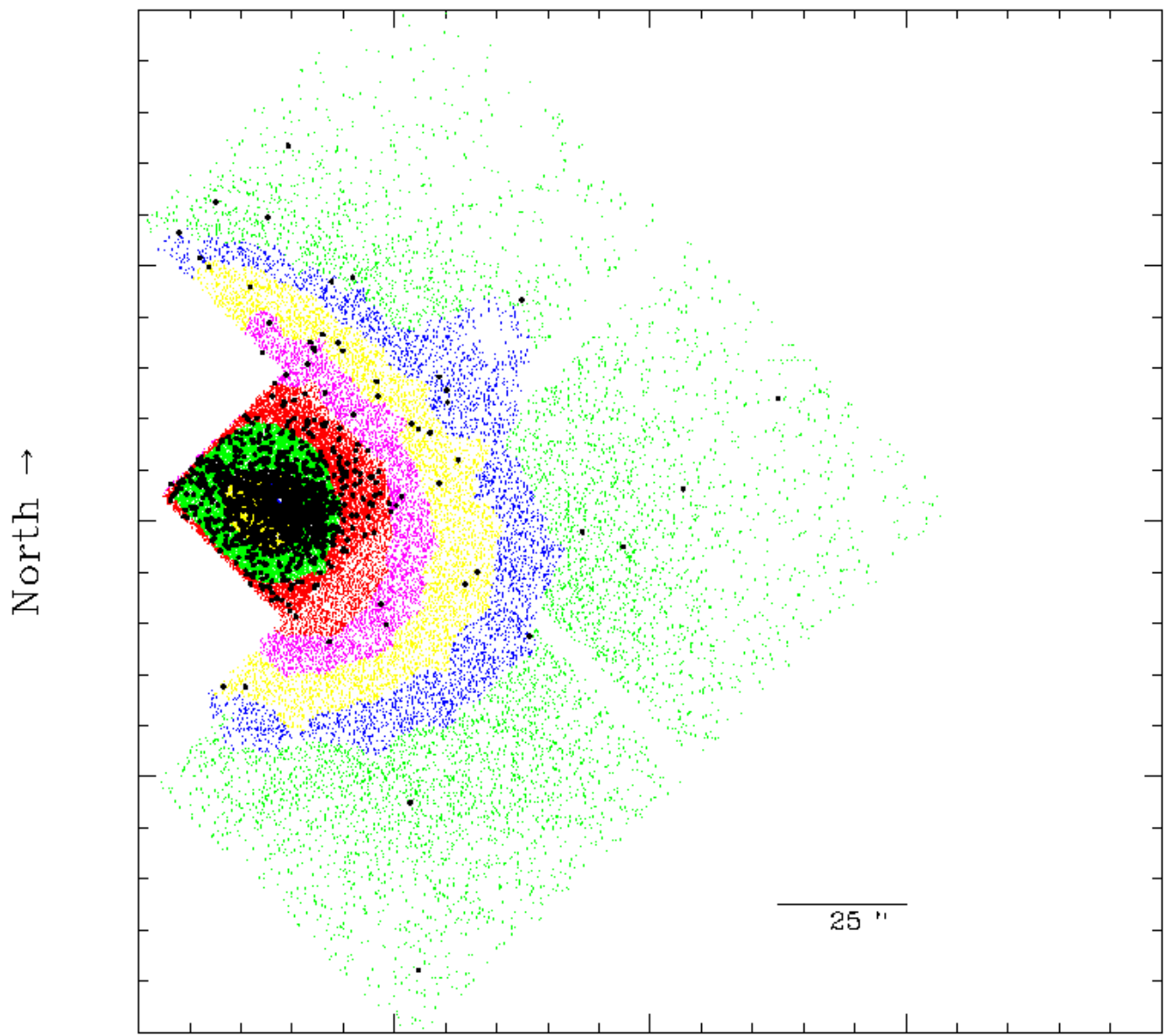}
\label{map}
\caption{31,351 stars measured with 
 WFPC2 in both F555W and F814W filters (dots) (T01). Black dots 
 correspond to stars with \mvi$<0$ and \mi$>22.5$.}
\end{figure}

The recent SFH of NGC~1705 is characterized by two bursts. 
The older one (B1), confined to the most central regions, is
concurrent with the strong activity that occurred from 15 to 10 Myr ago. Is
it an example of SF triggered by supernova explosions and stellar winds ? Or
a consequence of the SSC evaporation ? Do B1 and the SSC start simultaneously
or one is the cause of the other ? Larsen (2002) has recently suggested 
that SSCs form only in regions
of extremely high SF activity. However, the SF rate during B1 is not specially
high ($\sim$ 0.07 \MSUN $yr^{-1}$ for a Salpeter's IMF). A stronger young
burst (B2) 
started $\sim$ 3 Myr ago and is still ongoing with a total SFR of $\sim$0.3 
\MSUN $yr^{-1}$. Does this imply, in Larsen's hypothesis, that more SSCs
are currently being formed in NGC~1705 ?
The SF rate of B2 is much higher than those generally found for 
other BCDs  ($10^{-2}-10^{-3}$ \MSUN $yr^{-1}$). It is comparable only to the 
strong burst in NGC~1569 which took place from 100 Myr to 5-10 Myr ago at a 
rate of 0.5 \MSUN $yr^{-1}$ (Greggio et al. 1998). NGC~1569 contains indeed
three SSCs and several other young rich clusters (e.g. Hunter et al. 2000, 
Origlia et al. 2001). 
Such a young burst is not present in the bright SSC of NGC~1705, where the lack of spectral features from O and WR stars (Heckman \& Leitherer 1997) indicates a termination of the SF  at least 5-6 Myr ago. Evidence of very young stars in the high surface brightness region (HSB) was previously found by MFDC who detected objects significantly bluer than the SSC. Some of the objects are ionizing clusters in HII regions with ages less than $\sim$ 5 Myr. They found further evidence of very recent star formation in some broad emission features that they attributed to WR stars.  To compare their results with ours, we show in Fig.~\ref{map} the spatial distribution of our stars with age $\leq$3 Myr, overplotted on the galaxy isophotal map of T01's Fig.~1. Our youngest stars have locations very similar to those of MFDC's young embedded objects (their Fig.~3).  Particularly interesting is the concurrence of the north-east {\it finger} of our stars with theirs.
 Different from B1, which is traced only in the innermost regions, burst B2 has started over a larger galactic area. Its intensity decreases from the center: the highest contribution (0.27 \MSUN $yr^{-1}$) comes from Regions 7 and 6 (within 0.2 kpc from the center of the galaxy). In Regions 3 and 4 (between  $\sim$ 0.4 and $\sim$ 0.7 kpc from the center) the rate drops to $\sim 10^{-2}$ \MSUN $yr^{-1}$, and at distances $\gsim 1$ kpc we find only few stars younger than 3 Myr. The presence of extremely young stars even at large distances from the center of the galaxy could be related to the very extended H$\alpha$ surface brightness profile (Papaderos et al. 2002). 

What triggered this strong recent burst? MFDC proposed the SSC of NGC~1705 as the powering source for the spectacular bipolar wind observed. In their model, a hot bubble generated by supernova ejecta and stellar winds expands preferably along the minor axis, producing the bipolar wind. In contrast, the hot bubble compresses the neutral and molecular gas in the disk, inducing star formation and producing the observed HII regions.  Such a mechanism could have triggered the strong SF event that began just $\sim$ 3 Myr ago. On the other hand, three-dimensional numerical simulations  (Mori, Ferrara, \& Madau 2002) show that bubbles driven by off-center SN explosions in $\sim 10^8$ \MSUN \ halos tend to pile up cold gas and induce star formation in the central regions, while  more concentrated SN explosions produce an expanding mass flow.  Therefore, if B1 occurred in the center and was responsible for the expanding superbubble, we would expect to trace B2 (mostly) in the outer regions. On the contrary, our simulations yield the highest SF rates in Regions 7 and 6.

We can estimate the number of SN~II explosions over the past 50~Myr. The episode (3-0) Myr does not significantly contribute to the number of SNe~II. Hence, only stars formed during, or prior to the (15-10) Myr episode can have already exploded. We derive about 430 SNe II for B1 if we consider 30 \MSUN as upper mass cut-off for these events. The number of SNe almost doubles by increasing the upper mass limit to 120 \MSUN, releasing a total kinetic energy in the range 8$(10^{52}-10^{53})$ erg. Lisenfeld, Volk, \& Xu (1996) propose that these SNe haven't had time yet to produce any significant synchrotron radio emission due to strong inverse Compton losses. Therefore, the radio emission from the two most recent SF episodes is expected to be completely thermal, unless the magnetic field is rapidly increased (e.g., by SN shocks or a galactic wind).

The total astrated mass in stars younger than 1 Gyr is about $6 \times 10^{7}$ \MSUN, and that in stars older than 1 Gyr is at least $2.2 \times 10^{8}$ \MSUN. These values are moderately larger than those derived by MFDC, which are based on assumed M/L ratios. Given the different modeling approach, we consider our results in agreement with theirs. If NGC~1705 started forming stars $\sim$ 5 Gyr ago, our data are consistent with an approximately constant build up of the galaxy over its whole life, at an average rate of $\approx$ 0.06 \MSUN/yr. 

 SFRs are often expressed in terms of rate per unit area. We have thus calculated the SF rate densities of our episodes for a more direct comparison with other galaxies. In the case of NGC~1569, Greggio et al. (1998) derived a SFR of 4 \MSUN $yr^{-1} kpc^{-2}$ over an area of 0.14 $kpc^2$. The same area in NGC~1705 roughly corresponds to the sum of Regions 7 and 6, which give the major contribution (0.27 \MSUN $yr^{-1}$) to the burst B2, 3--0 Myr ago. In this area the SF density of burst B2 becomes $\sim$ 2 \MSUN $yr^{-1} kpc^{-2}$, close to that of NGC~1569. However, we know that B2 is occurring not only in the central regions but almost everywhere. If we consider the whole galactic area sampled (12.5 $kpc^2$), the SF density of this burst drops to $\sim$ 0.024  \MSUN $yr^{-1} kpc^{-2}$, much lower than that of NGC~1569 and close to the range of  those derived both in Local Group irregulars  ($10^{-4}-10^{-2}$ \MSUN $yr^{-1} kpc^{-2}$; e.g., Tosi 1999) and in other BCDs (IZw18: 0.01--0.1 \MSUN $yr^{-1} kpc^{-2}$, Aloisi et al. 1999; UGCA 290: 0.01 \MSUN $yr^{-1} kpc^{-2}$, Crone at al. 2002; Mrk 178: (1 $\div$ 5) $10^{-2}$ \MSUN $yr^{-1} kpc^{-2}$, Schulte-Ladbeck et al. 2000; VIIZw 403: 0.02 \MSUN $yr^{-1} kpc^{-2}$, Schulte-Ladbeck et al. 1999b, Crone et al. 2002).  Burst B1 (15--10 Myr ago) is one order of magnitude less intense than B2: if considered over the area (0.14 $kpc^2$) studied in NGC~1569, the SF density translates into 0.5 \MSUN $yr^{-1} kpc^{-2}$. This is lower than the burst of NGC~1569 but relatively high compared to the SF density in other irregulars and BCDs.  However, when calculated over the entire region where stars generated by B1 are found ($\sim 0.5 kpc^2$), the rate density of B1 is reduced to 0.14 \MSUN $yr^{-1} kpc^{-2}$.

 At intermediate epochs, the average SFR is not particularly strong ($\sim$ 0.056 \MSUN $yr^{-1}$) but still larger than found for other nearby BCDs (of the order of $10^{-3}-10^{-2}$ \MSUN $yr^{-1}$, e.g., see the reviews in Schulte-Ladbeck 2001b, or in Tosi 2001). When normalized to the area where most of the intermediate age population is seen ($\sim 1.35 \ kpc^2$), the rate density becomes 0.04 \MSUN $yr^{-1} kpc^{-2}$, i.e., close to the value observed in other BCDs.

 The oldest episode 5--1 Gyr ago occurred in the whole area covered by the PC and the three WFs ($\sim 12.5 kpc^2$), with an average rate of  $4.5 \ 10^{-3}$ \MSUN $yr^{-1} kpc^{-2}$. This is comparable to the values derived for Local Group irregulars. 

%\subsection {The IMF}

We adopt a single IMF slope from the upper to the lower mass limit. However, the IMF mostly affects the distribution of the youngest stars, and the slope is constrained only for M $\gsim$ 6 \MSUN. A Salpeter or slightly steeper IMF  ($\alpha=2.6$) gives the best match to the data. This result is universally found for almost all the other dwarfs studied with the method of the synthetic CMDs. The only case with evidence for a flat IMF ($\alpha=1.5-2$) is IZw18 (Aloisi et al. 1999). Our result on the IMF of NGC~1705 is in contrast with previous preliminary work, where the synthetic CMDs suggested a rather flat IMF ($\alpha \sim 1.5$, Annibali et al. 2001). There are two main causes for this disagreement. The first is an incorrect treatment of the incompleteness in our previous work, as we did not account for the different crowding conditions of regions at different distances from the galaxy center. Variations in the computed incompleteness factors affect the ratio of faint to bright stars in the synthetic CMDs, and can then affect the derived IMF slope. A second cause is our initial choice of using only the objects whose photometry is available in both the optical (F555W and F814W) and NIR (F110W and F160W) bands. The NIR turns out to be almost blind to the episode (2-0) Myr: this extremely young burst populates the optical CMDs with some hundreds of blue and relatively faint stars. These stars are hardly recoverable in the NIR plane, where our photometry is much shallower than in the optical. Consequently, by choosing a subsample of stars detected both in the optical and in the NIR the most recent SF activity is lost. A low SF at recent times, combined with a flat IMF, was sufficient to overpopulate the blue plume. In contrast, the simulations presented here are based on the whole optical sample and demonstrate that the difficulties in populating the blue plume are not a matter of the IMF but of the SFH. An additional young SF episode is required.  

Are these results compatible with the BF model for the interpretation of the faint galaxy counts? We cannot reliably measure the duration of the old SF episodes. We tested a simulation in which a burst of star formation occurred 5 Gyr ago, at a rate of 1 \MSUN $yr^{-1}$, and lasting 100 Myr. This burst produced only about half of the total mass in the old component, leaving room for further evolution. In spite of the short duration, the RGB of this simulation turned out smooth and wide, because of the effect of the photometric error. Although the SF history over the most recent 1 Gyr appears rather continuous, the early conditions could have favored a more bursty SF history. The strength of the current burst B2 is indeed not far from what required in the BF model. Therefore, NGC~1705 could have been one of the bursting dwarfs at high redshift. However, different from the bursting dwarfs in BF model, further evolution did certainly occur in this galaxy, which appears fairly active at the current epoch. In the case of Mrk 178, Schulte-Ladbeck et al. (2000) found that the galaxy should have formed all the old stars in an extremely short period at early times and then remained inactive, in order to have the SFR required by the BF model.

In NGC~1705, the total gas mass, including the neutral and ionized hydrogen, and corrected for the helium content, is comparable to the stellar mass ($M_g \sim 1.2 \ 10^8$ \MSUN, MDFC). We can calculate the time in which all the present gas mass would be used up by future star formation at a given rate. If we adopt a rate of 0.06 \MSUN $yr^{-1}$, derived for the SF activity earlier than 1 Gyr ago,  we obtain that the SF can be maintained for another 2 Gyr.  With an IMF that flattens at low masses ($\alpha=-0.564$ in $0.1<$\MSUN$<0.6$, Gould, Bahcall, \& Flynn 1997), the SFR decreases to $\sim 0.032$ \MSUN $yr^{-1}$, and the gas consumption time becomes 3.7 Gyr. If instead we adopt the current rate of $\sim 0.3$ \MSUN $yr^{-1}$ ($\sim 0.16$ \MSUN $yr^{-1}$ for the IMF of Gould et al.), the consumption time for the present gas mass is $\sim 400$ Myr (750 Myr). From the actual gas mass and the total stellar mass astrated since the earliest epochs ($3 \ 10^8$ \MSUN for a Salpeter slope down to 0.1 \MSUN, $1.6 \ 10^8$ \MSUN for the same IMF), we obtain that the initial gas mass ($4.2 \ 10^8$ \MSUN \ or  $2.8 \ 10^8$ \MSUN) is 2-4 times larger than the present day gas mass.  

Our major results are the following: 

\begin{itemize}

\item{ NGC~1705 is not a young galaxy but has started forming stars  several Gyrs ago. There is some evidence for a more metal-poor old population in the most central regions than in the periphery.} 
\item{ While the old population is spread throughout the entire galaxy,   the young and intermediate-age stars  are mostly located in the central regions.} 
\item{ The derived SF is almost continuous with fluctuations in the SFR. Rather than a bursting SF history, characterized by short and intense episodes with intermittent quiescent periods, the most likely history is {\it gasping}, where episodes of relatively high activity are separated by less active phases.  The rates are not especially high and comparable to those found for other BCDs. The only exception is the episode starting $\sim $3 Myr ago and still continuing with a total rate of 0.3 \MSUN $yr^{-1}$.} 
\item{We find a clear signature of a burst SF in the field. This burst occurred from 15 to 10 Myr, almost simultaneously with the strong SF which generated the SSC.   The mass in field stars in this episode is larger than that in the SSC.}
 \item{Our estimate of the total stellar mass in this galaxy is $\gsim$ 2.8 $\times 10^{8}$, with $\sim$ 21 \%   in stars younger than 1 Gyr.} \item{A Salpeter or slightly steeper ($\alpha=2.6$) IMF gives the best agreement with the observations.} \end{itemize}

\acknowledgments

We thank Matteo Monelli, Paolo Montegriffo and Elena Sabbi for their help to properly handle the photometric data. Interesting conversations with A. D'Ercole, F.R. Ferraro, L. Gregorini, U. Hopp and G. Zamorani are gratefully acknowledged. This work has been partly supported by the Italian ASI, through grants ARS-99-44 and ASI--I/R/35/00, and by the Italian MIUR, through Cofin2000. Part of the work of F.A. was presented as Laurea Thesis at the Bologna University.

\clearpage
               
\begin{deluxetable}{llll}
%\tabletypesize{\small}
\tablewidth{0pt}
\tablenum{1}
\tablecaption{Region 7\label{tab_1}}
\tablecolumns{4}
\tablehead{
\multicolumn{2}{c}{Episode (Myr ago)} &  \multicolumn{2}{c}{SFR 
 (\MSUN $yr^{-1}$)}  \\
\\[0pc]
\colhead{Start} & \colhead{End} & \colhead{$\alpha=2.35$} & 
\colhead{$\alpha=2.$} 
}
\startdata
 2-3   &   0  & $2.15^{\mp 0.55} \times 10^{-1}$ &  
 $6.5^{\mp 1.7} \times 10^{-2}$ \\ 
%& $2.70^{\mp 0.66} \times 10^{-2}$ \\
\\[0pc]
15-16   & 9-10  & $4.25^{\mp 0.15} \times 10^{-2}$ 
& $1.85^{\mp 0.25} \times 10^{-2}$ \\ 
%& $1.35^{\mp 0.15} \times 10^{-2}$ \\
\\[0pc]
100-600  &  40-60 & $7.8^{\mp 3.6} \times 10^{-3}$ 
& $5.3^{\mp 2.4} \times 10^{-3}$ \\ 
%& $5.8^{\mp 2.4} \times 10^{-3}$ \\
\enddata
\tablecomments{For all the regions, 
 the SFRs have been derived assuming a constant IMF slope from 
 0.1 \MSUN $<M<$ 120 \MSUN.} 
\end{deluxetable}

\clearpage
               
\begin{deluxetable}{lllll}
%\tabletypesize{\small}
\tablewidth{0pt}
\tablenum{2}
\tablecaption{Region 6\label{tab_2}}
\tablecolumns{5}
\tablehead{
\multicolumn{2}{c}{Episode (Myr ago)} &  \multicolumn{2}{c}{SFR 
 (\MSUN $yr^{-1}$)} & \colhead{Z} \\
\\[0pc]  
\colhead{Start} & \colhead{End} & \colhead{$\alpha=2.6$} & 
\colhead{$\alpha=2.35$}& \colhead{} }
\startdata
 2 - 3  & 0  & $(5.6 - 3.4) \times 10^{-1}$ & $(1.7 - 1.1) \times 10^{-1}$ &
 0.004 \\
\\[0pc]
16 - 17   &   10 - 11  &  $(5.2 - 2.8) \times 10^{-2}$
& $(2.3 - 1.6) \times 10^{-2}$ &  0.004  \\
\\[0pc]
50 - 60   &   16 - 17   & $(1.3 - 0.9) \times 10^{-2}$
& $(7.5 - 5.2) \times 10^{-3}$ & 0.004  \\
\\[0pc]
100       &   50 - 60   &  $(5.8 - 4.0) \times 10^{-2}$
& $(3.5 - 2.5)  \times 10^{-2}$  &  0.004 \\
\\[0pc]
 $1 \times 10^{3}$  & 100  & $(4.8 - 2.2) \ 10^{-2}$ &
$(2.8 - 1.9) \times 10^{-2}$ &  0.004 \\
\\[0pc]
$ (2 - 15) \times 10^{3}$   &  $1 \times 10^{3}$
  & $(5.3 - 0.3) \times 10^{-2}$  & $(3.5 - 0.5) \times 10^{-2}$ & 0.0004 \\
\enddata

\end{deluxetable}

\clearpage

\begin{deluxetable}{lllll}
%\tabletypesize{\small}
\tablewidth{0pt}
\tablenum{3}
\tablecaption{Region 5\label{tab_3}}
\tablecolumns{5}
\tablehead{
\multicolumn{2}{c}{Episode (Myr ago)} 
& \multicolumn{2}{c}{SFR (\MSUN $yr^{-1}$)} 
& \multicolumn{1}{c}{Z} \\
\\[0pc]  
\colhead{Start} & \colhead{End} & \colhead{$\alpha=2.6$} & 
\colhead{$\alpha=2.35$} & \colhead{} 
}
\startdata
 2-3   &   0    &  $(14 - 9.3) \times 10^{-2}$ & $(5.3 - 3.2) \times 10^{-2}$ 
 & 0.004 \\
\\[0pc]
 40-50 & 10-12  &  $(7.9 - 3.9) \times 10^{-3}$ & $(4.2 - 2.5) \times 10^{-3}$
 & 0.004 \\
\\[0pc]
 300   & 40-60  &  $(3.1 - 1.6) \times 10^{-2}$ & $(1.8 - 1.1) \times 10^{-2}$ 
 & 0.004 \\
\\[0pc]
 500   &  300   &  $(3.1 - 1.8) \times 10^{-2}$ & $(2.1 - 1.2) \times 10^{-2}$ 
 & 0.004\\
\\[0pc]
 $1 \times 10^{3}$ &   500   &  $(6.6 - 1.8) \times 10^{-2}$ & 
 $(4.2 - 1.2) \times 10^{-2}$ 
 & 0.004 \\
\\[0pc]
 $(10 - 15) \times 10^{3}$  &  $1 \times 10^{3}$   & $(10.0 - 4.4)
 \times 10^{-3}$ 
 & $(8.3 - 4.3) \times 10^{-3}$ & 0.0004 \\
 % &  (3-1) Gyr & $1.4 \times 10^{-2}$ &  $ 1.2 \times 10^{-2}$ & 0.004 \\
%\\[0pc]
%  &  (5-3) Gyr   & $1.7 \times 10^{-2}$ & $1.5 \times 10^{-2}$ & 0.0004\\
\enddata

\end{deluxetable}

\clearpage

\begin{deluxetable}{llll}
%\tabletypesize{\small}
\tablewidth{0pt}
\tablenum{4}
\tablecaption{Regions 3-4\label{tab_4}}
\tablecolumns{4}
\tablehead{
\multicolumn{2}{c}{Episode (Myr ago)} & 
\colhead{SFR (\MSUN $yr^{-1}$)} & 
\colhead{Z} \\
\\[0pc]
\colhead{Start} & \colhead{End} &\colhead{} & \colhead{} 
} 
\startdata
  2-3       &   0     & $(11 -7) \times 10^{-3}$  &    0.004   \\
\\[0pc]
  300        & 50-70     & $(1.2 - 1.1) \times 10^{-3}$  &    0.004 \\
\\[0pc]
 $1 \times 10^{3}$   & 300   & $\leq 5.0 \times 10^{-2}$  &  
0.004 \\
%\tableline
%\\[0pc]
\\[0pc]  
$3 \times 10^{3}$  & $1 \times 10^{3}$  & $(2.6 - 1.9) \times 10^{-2}$  
& 0.004 \\
%\tableline
\\[0pc]
%\\[0pc]
% $5 \ 10^{3}$   &$1\ 10^{3}$ &  $6 \ 10^{-3}$ & 0.004\\
% $10 \ 10^{3}$   &$5\ 10^{3}$ &  $2.6 \ 10^{-3}$ & 0.0004\\
\bf{Z-mixed} & & & \\
%\\[0pc]
$3 \times 10^{3}$  & $1 \times 10^{3}$  & $\sim 1.9 \times 10^{-2}$  
& 0.004 \\
$15 \times 10^{3}$  & $3 \times 10^{3}$  & $\sim 2.2 \times 10^{-3}$  
& 0.0004 \\
\\[0pc]
\enddata

\end{deluxetable}

\clearpage

\begin{deluxetable}{lll}
%\tabletypesize{\small}

\tablewidth{0pt}
\tablenum{5}
\tablecaption{Regions 0-1-2\label{tab_5}}
\tablecolumns{3}
\tablehead{
\multicolumn{2}{c}{Episode (Myr ago)} &  \colhead{SFR (\MSUN $yr^{-1}$)} \\
\\[0pc]
\colhead{Start} & \colhead{End}  & \colhead{} \\
}
\startdata
 2-20  &  0     &  $(5.2 - 0.3) \times 10^{-3}$ \\
\\[0pc]
 1000  & 50 -70 &  $(6.2 - 5.0) \times 10^{-4}$ \\
\\[0pc]
$(3-5) \times 10^{3} $  & $ 1 \times 10^{3}$ & $(2.6 - 1.7)\times 10^{-2}$ \\
\\[0pc]
\enddata

\end{deluxetable}

\clearpage

\begin{deluxetable}{llllllll}
\rotate
\tablewidth{0pt}
\tablenum{6}
\tablecaption{Average SFRs at various epochs \label{tabsummary}}
\tablecolumns{8}
\tablehead{
\colhead{Region} & \colhead{Area ($kpc^2$)} & \multicolumn{6}{c}{SFR (\MSUN $yr^{-1}$)} \\
\\[0pc]
\colhead{} & \colhead{} & \colhead{(0-3) Myr} & 
\colhead{(3-10) Myr } & \colhead{(10-15) Myr} & \colhead{(15-50) Myr}
& \colhead{(50-1000) Myr} & \colhead{(1-5) Gyr}\\
} 
\startdata
 7 & 0.017& 0.16 & $-$ & 0.044 &
 $-$  & $4.3 \times 10^{-3}$ & ? \\
\\[0pc]
 6      & $0.13$ & 0.11 & $-$ & 0.022 & 
 0.005 & 0.024 & $1.3 \times 10^{-2}$ \\
\\[0pc]
 5      & $0.35$     & 0.032 & $-$ & 
 $2.1 \times 10^{-3}$ & $2.7 \times 10^{-3}$ & 0.022 &
 $1.6 \times 10^{-2}$ \\
\\[0pc]
 3--4   & $0.85$   & 0.007 & $-$ & $-$ & $-$ 
 &$7.4 \times 10^{-3}$ 
 & $1.0 \times 10^{-2}$ \\
\\[0pc]
0--1--2 & $11.2$   & 0.005 & $-$ & 
$-$ & $-$ & $5.0 \times 10^{-4}$ 
&$1.7 \times 10^{-2}$ \\
\tableline
\\[0pc]
total  &  12.5 &  0.314   &  $-$ & $6.8 \times 10^{-2}$  &  
$7.7 \times 10^{-3}$ & $5.8 \times 10^{-2}$  & $5.6 \times 10^{-2}$  \\
\enddata

\end{deluxetable}

\clearpage

\clearpage

\clearpage

\clearpage

\clearpage

\clearpage

\clearpage

\clearpage

\clearpage

\clearpage

\clearpage

\clearpage

\clearpage

\clearpage

\end{document}